\documentclass[showpacs,amsmath,amssymb,prl,superscriptaddress,floatfix]{revtex4}

\usepackage{graphicx}
\usepackage{dcolumn}
\usepackage{bm}
\newcommand{\be}{\begin{equation}}
\newcommand{\ee}{\end{equation}}

\newcommand{\beq}{\begin{eqnarray}}
\newcommand{\eeq}{\end{eqnarray}}

\def\H1{\widehat{H}_1}

\begin{document}


\title{Semiclassical solitons in strongly correlated systems of ultracold 
bosonic atoms in optical lattices}
\date{\today}
\pacs{03.75.Be, 32.80.Pj, 42.50.Vk}

\author{Eugene Demler}
\affiliation{Department of Physics, Harvard University, Cambridge MA
02138, USA}
\author{Andrei Maltsev}
\affiliation{L.D.Landau Institute for Theoretical Physics,
Chernogolovka, Moscow reg. 142432, Russia }

\date{\today}

\begin{abstract}
We   investigate  theoretically 
soliton excitations and dynamics of their formation in strongly 
correlated  systems of ultracold bosonic atoms in two and three dimensional
optical lattices. We  derive equations of
nonlinear hydrodynamics in the regime of strong interactions and incommensurate fillings, 
when atoms can be treated as hard core bosons. When parameters 
change in one direction only we obtain Korteweg-de Vries type 
equation away from half-filling and modified KdV equation
at half-filling.
We apply this general analysis to a problem of the decay of the 
density step. We consider stability of one dimensional solutions to 
transverse fluctuations. Our results are also relevant for understanding 
nonequilibrium dynamics of lattice spin models.
\end{abstract}

\maketitle

\section{ Introduction}

Solitons are conspicuous manifestations of nonlinear interactions in a 
variety of physical systems (see e.g. \cite{Ostrovsky1999,Kivshar1989}).
Originally introduced in hydrodynamics of classical fluids, they were 
later observed in a variety of other systems, including plasma physics, 
nonlinear optics, magnetism, dynamics  of molecular systems. It is 
currently understood that formation of oscillatory zones and localized 
solitonic solutions  is a common feature of many non-linear systems and 
does not depend on the exact integrability of the model. However the 
character of solitons is different for each system and understanding 
their properties remains a fundamental problem in physics and 
mathematics.

In this paper we   investigate  theoretically the nature of 
solitons and dynamics of their formation in strongly correlated  systems
of ultracold bosonic atoms in optical 
lattices \cite{Jaksch2005,Bloch2005,Lewenstein2007,Bloch2008_1}. Recently 
questions of far from equilibrium many-body dynamics took central stage 
in both theoretical and experimental study of ultracold atoms. What makes 
such systems particularly well suited for exploring quantum dynamics is 
their good isolation from the environment. Their characteristic energies 
and frequencies are of the order of kiloHertz, which is extremely 
convenient for experimental studies. It is also important that a wide array of experimental tools 
that allow to control system parameters in time and prepare far from 
equilibrium initial states have been developed. Recent experiments 
addressed such question as dynamics of fermions in optical
lattices \cite{Strohmaier2010,Schneider2010}, observation of  
superexchange interactions using spin dynamics \cite{Trotzky2008}, 
thermalization and relaxation in one-dimensional
systems \cite{Kinoshita2006,Hofferberth2007,Haller2009}, 
motion of impurity particles \cite{Kohl2009}, dynamics
and adiabaticity in crossing classical and quantum phase
transitions \cite{Ritter2007,Sadler2006}. Another important 
recent achievement is development of experimental tools for the in-situ 
imaging of individual atoms in optical 
lattices 
\cite{Nelson2007,Ott2008,Gemelke2009,Karski2009,Bakr2010,Kuhr2010} 
and low dimensional condensates \cite{Schmiedmayer2010,Moritz2010}. This 
technique allows unprecedented level of characterization of many-body
states and should lead to deeper understanding of their out of 
equilibrium dynamics. One example is recent analysis of Bakr et al 
\cite{Bakr2010} of the dynamics of defect creation in crossing from the SF 
to Mott state in two dimensional optical lattices.

We start our analysis by deriving hydrodynamical approach to describe 
quantum dynamics of the lattice bosons. Hydrodynamical description has 
been applied to quantum many-body systems previously, including
superfluids \cite{Khalatnikov1978}, superconductors \cite{Greiter1989}, 
quantum Hall systems \cite{Wen1995}, and magnets\cite{Halperin1969}. 
The focus of most earlier analysis was on understanding
collective modes and universal features of linear response functions, 
which only required understanding linear hydrodynamics.  When non-linear
effects have been discussed for superfluid systems, it was primarily done
for systems in the continuum with the the full Galilean symmetry. Our goal will be
to include both nonlinearities and dispersion, since the competition of 
the two determines the nature and dynamics of solitons.

Solitons in systems of ultracold atoms have been discussed
previously in the regimes where semiclassical Gross-Pitaevskii equation can 
be applied either in uniform systems 
\cite{Burger1999,Denschlag2000,Khaykovich2002,Pitaevskii2003,Castin2009}
or systems with optical lattices
\cite{Eiermann2004,Trombettoni2001,Ahufinger2004,Ahufinger2005,Yulin2003,Johansson1999,Kevrekidis2003}. 
In this paper we will be
interested in the regime of very strong interactions between
atoms, the so-called hard core bosons regime 
\cite{Scalettar1995,Schmid2002}. 
In this case dynamics of atoms in a lattice 
can be described using anisotropic Heisenberg model\cite{Scalettar1995}.
We demonstrate that in this regime the character of soliton
excitations is very different and depends on both the filling factor
and parameters of the Heisenberg model. 
Numerical analysis of solitons in Bose systems in optical lattices
in the vicinity of the SF/Mott transition has been done 
recently by Krutitsky et al. \cite{Krutitsky2010}.
Our results can also be applied to study nonequilibrium spin dynamics of
two component Bose mixtures in the Mott
insulating regime\cite{Duan2003,Kuklov2003} and lattice spin systems
in solid state physics.


\section{Model}

\subsection{From lattice bosons to spin Hamiltonian}

Microscopic model describing ultracold bosonic atoms  in an optical 
lattice is given by the Bose-Hubbard model\cite{Jaksch2005,Bloch2008_1}
\begin{eqnarray}
{\cal H}_{\rm BH} = -t_{\rm b} \sum_{\langle ij \rangle } b_i^\dagger b_j 
+ \frac{U}{2} \sum_i n_i ( n_i - 1) 
\label{hubbard}
\end{eqnarray}
Here $b_i^\dagger$ is a creation operator for bosons on site $i$, 
$n_i=b_i^\dagger b_i$ is the number of atoms on site $i$.
We do not include the chemical potential term because in this paper we  
study dynamics and the operator of the total number of particles commutes 
with the Hamiltonian. It is sufficient to impose a certain number of 
particles at the initial time and then the total number of particles 
should not change during evolution.
When there is inhomogeneous external potential we also need to add
\begin{eqnarray}
V_{\rm ext} = \sum_{i}  V_{i}   n_{i} 
\end{eqnarray}

To keep the model more general we  include nearest neighbor interactions
\begin{eqnarray}
{\cal H}_{\rm Ext \, BH} = {\cal H}_{\rm Hub} + V \sum_{\langle ij 
\rangle} n_i n_j
\label{hubbard_extended}
\end{eqnarray}
Such non-local interaction are relevant for atoms in higher Bloch 
bands\cite{Scarola2005} and polar molecules in optical 
lattices\cite{Lahaye2009}. We consider a regime when the local repulsion 
$U$ is large and the density of particles is incommensurate with the 
lattice. In this case strong number fluctuations are suppressed even in 
the superfluid state and we can limit the  Hilbert space  to only two  
possible occupation numbers $| n_0-1 \rangle$ and $| n_0 \rangle$ per 
site. It is convenient to  represent these states as spin states. State $| 
n_0-1 \rangle_i$ corresponds to $| \downarrow \rangle_i$, and state
$| n_0 \rangle_i$ corresponds to $| \uparrow \rangle_i$. In this limit 
Hamiltonian (\ref{hubbard_extended}) is equivalent to the
anisotropic Heisenberg model
\begin{eqnarray}
{\cal H}_{\rm AH} = - J_\perp \sum_{\langle ij \rangle}  \left( 
\sigma_i^x \sigma_j^x + \sigma_i^y \sigma_j^y \right)
- J_z \sum_{\langle ij \rangle} \sigma_i^z \sigma_j^z 
\label{heisenberg}
\end{eqnarray}
Here $\sigma^a$ are Pauli matrices, $2 J_\perp = t_{\rm b} \, n_0$,  and 
$J_z = -V$.

Hamiltonian (\ref{heisenberg}) also appears as an effective description 
of spin dynamics in the Mott state of two component Bose mixtures at 
filling factor $n=1$\cite{Duan2003,Kuklov2003}.

\subsection{Semiclassical equations of motion for lattice bosons}

In this section we discuss how one can obtain semiclassical description 
of dynamics of (\ref{heisenberg}) using either variational Gutzwiller 
wavefunctions or linearized equations of motion, which in this case
are equivalent to lattice Landau-Lifshitz equations. To simplify the
derivation we assume that parameters of the system change  in one 
direction only. We emphasize that our focus is on two and three 
dimensional systems. Restriction to having variations
of parameters in only one direction is, firstly, for notational 
simplicity (extension to higher dimensions is straightforward)
and, secondly, because we will be concerned with problems where initial 
state has been prepared to have
parameters changing along one of the coordinates. We discuss effects of 
fluctuations in transverse directions in subsequent sections.  

Strictly one dimensional systems are special and
mean-field approaches do not apply to them even in equilibrium. However 
special analytical approaches are available for one dimensional systems, 
including fermionization and Bethe 
ansatz\cite{Fradkin1991,Giamarchi2004,Sutherland2004}.
Also powerful numerical methods based on DMRG \cite{Schollwock2005} and 
Matrix Product States\cite{Daley2004} allow to study dynamics of one 
dimensional systems in great details. On the other hand, nonequilibrium 
dynamics of higher dimensional systems remains
largely unexplored. This is the main motivation for the current paper. 
Interestingly, recent work by Lancaster and Mitra \cite{Lancaster2010} 
showed that semiclassical analysis of Landau-Lifshitz equations for one 
dimensional spin chains give results consistent  with exact calculations. 
Hence our results may also be relevant for one dimensional spin chains.

To obtain semiclassical dynamics we consider time-dependent variational 
wavefunctions
\begin{equation}
\label{WaveFunction}
|\Psi (t) \rangle \, = \, \prod_{i} \left[ \sin {\theta_{i} (t)\over 2}
\,\, e^{-i \varphi_{i}(t)/2} \, |\downarrow \rangle_{i} \, + \,
\cos {\theta_{i}(t) \over 2}  \,\, e^{i \varphi_{i}(t)/2} \,
|\uparrow \rangle_{i} \right]
\end{equation}
Expectation values of the original boson operators are 
\begin{eqnarray}
\langle b_i \rangle &=& \frac{\sqrt{n_0}}{2} 
\sin \theta_i e^{-i \varphi_i} 
\nonumber \\
\langle n_i \rangle & = & n_0 + \frac{1}{2} ( \cos \theta_i -1) 
\end{eqnarray}

 To project Schrodinger equation into wavefunction (\ref{WaveFunction}) 
we define the Lagrangian
\cite{Jackiw1979,Huber2008}
$$L \,\, = \,\, - i \, \langle \Psi | \frac{ d}{ dt} | \Psi \rangle 
\, + \, \langle \Psi | H | \Psi \rangle \,\, = $$
$$= \,\, \sum_{i} {1 \over 2} \, {\dot \varphi}_{i} \,
\cos \theta_{i} \,\, - \,\, J_{\perp} \, \sum_{<ij>}
\sin \theta_{i} \, \sin \theta_{j} \,
\cos (\varphi_{i} - \varphi_{j}) \,\, - \,\,
J_{z} \, \sum_{<ij>} \cos \theta_{i} \, \cos \theta_{j} $$
and write equations of motion
\begin{eqnarray}
\frac{d}{dt} \frac{\delta L}{\delta \dot{q}_i} -  
\frac{\delta L}{\delta {q}_i} = 0
\end{eqnarray}
Here $q_i$ corresponds to both $\varphi_i$ and $\theta_i$. We find
\begin{equation}
\label{DiscreteSystem}
\begin{array}{c}
{\dot \varphi}_{i} \, \sin \, \theta_{i} \,\, = \,\,
- \, 4 \, J_{\perp} \, \cos \theta_{i} \, \left( \sin \theta_{i+1} 
\, \cos (\varphi_{i} - \varphi_{i+1}) \, + \, \sin \theta_{i-1} \,
\cos (\varphi_{i} - \varphi_{i-1}) \right) \, + 
\cr
+ \, 4 \, J_{z} \, \sin \, \theta_{i} \, \left( \cos \theta_{i+1}
\, + \, \cos \theta_{i-1} \right)
\cr
\cr
{\dot \theta}_{i} \,\, = \,\, - \, 4 \, J_{\perp} \, \left(
\sin \theta_{i+1} \, \sin (\varphi_{i} - \varphi_{i+1}) \, + \,
\sin \theta_{i-1} \, \sin (\varphi_{i} - \varphi_{i-1}) \right)
\end{array}
\end{equation}
The first equation is effectively the Josephson relation: time 
derivative of the phase $\varphi$ is equal to the chemical potential 
which depends on the values of $\theta$ and $\varphi$. The second equation 
is charge conservation.

One can give an alternative physical interpretation to equations 
(\ref{DiscreteSystem}).
We write equations of motion for spin operators 
\begin{eqnarray}
\frac{d \sigma_i^x}{dt} = - 2 J_\perp \sigma_i^z (  \sigma_{i-1}^y + 
\sigma_{i+1}^y ) + 2 J_z \sigma_i^y ( \sigma_{i-1}^z + \sigma_{i+1}^z ) 
+ h^i_z \sigma_i^y
\end{eqnarray}
 And we have analogous equations for $\sigma_i^{\{y,z\}}$. 
To obtain semiclassical dynamics
we replace operators by their expectation values
\begin{eqnarray}
\frac{d \langle \sigma_i^x \rangle }{dt} = 
- 2 J_\perp \langle \sigma_i^z \rangle \left(  \langle \sigma_{i-1}^y 
\rangle + \langle \sigma_{i+1}^y \rangle \right)
+ 2 J_z \langle \sigma_i^y \rangle \left(  \langle \sigma_{i-1}^z \rangle 
+ \langle \sigma_{i+1}^z \rangle \right)
+ B^i_z \langle \sigma_i^y \rangle
\end{eqnarray}
 These are familiar Landau-Lifshitz equations. If we use wavefunction 
(\ref{WaveFunction}) to calculate $\langle \sigma_i^z \rangle = \cos 
\theta_i $ and 
$\langle \sigma_i^+ \rangle = \frac{1}{2} \sin \theta_i e^{-i\varphi_i}$, 
we recognize that Landau-Lifshitz equations are equivalent to 
(\ref{DiscreteSystem}).

Dynamics of the Bose-Hubbard model has been studied using Gutzwiller 
variational wavefunctions in 
\cite{Zakrzewski2005,Damski2003,Murg2007,Huber2007}.
In \cite{Altman2005,Polkovnikov2005} this approach was used 
to describe current decay in the strongly interacting regime of bosons. 
Theoretical predictions were in quantitative agreement with subsequent 
experimental results by Mun et al\cite{Mun2007}.

\section{Semiclassical dynamics in the continuum limit}

\subsection{Long wavelength expansion}

It is convenient to introduce slow variables in space, $X=h x$, and time, 
$T=ht$, where $h$ is the lattice constant. 
We are looking at dynamics of fluctuations that are slow on the scale of 
the lattice constant. So $h$ is a small parameter in which we will expand.
We introduce  
\begin{eqnarray}
\mu &=& \cos \theta
\nonumber\\
\sigma(X,T) &=& h \varphi(X,t).
\end{eqnarray}
and obtain
\begin{eqnarray}
{\cal L} \,\, = \,\, {1 \over 2} \, \sigma_{T} \, \mu \, - \,
2 \, J_{\perp} \, (1 - \mu^{2}) \, \cos \sigma_{X} \, - \,
2 \, J_{z} \, \mu^{2} \, + \, h^{2} \, J_{\perp} \, 
{\mu^{2} \mu_{X}^{2} \over 1 - \mu^{2}} \, \cos \sigma_{X} \, + 
\nonumber\\
\label{ApprLagr}
+ \, h^{2} \, J_{\perp} \, (1 - \mu^{2}) \,
\left( {1 \over 3} \, \sigma_{XXX} \, \sin \sigma_{X} \, + \,
{1 \over 4} \, \sigma_{XX}^{2} \, \cos \sigma_{X} \right) 
\, - \,  h^{2} \, J_{z} \, \mu \, \mu_{XX} \,\, + \,\, 
{\cal O} (h^{4}) 
\end{eqnarray}

\subsection{Hydrodynamics}

If we keep only the lowest order terms in $h$ in  (\ref{ApprLagr}),
we obtain the hydrodynamic part of the lagrangian 
\begin{eqnarray}
{\cal L}_{Hydr} \,\, = \,\, {1 \over 2} \, \sigma_{T} \, \mu \, - \,
2 \, J_{\perp} \, (1 - \mu^{2}) \, \cos \sigma_{X} \, - \,
2 \, J_{z} \, \mu^{2}
\label{lagrangian_hydrodynamic}
\end{eqnarray}

It is convenient to define
\begin{eqnarray}
k (X, T) \,\, = \,\, \sigma_{X} (X, T)
\nonumber\\
\end{eqnarray}
The new variable is proportional to the phase gradient, 
$ k \sim \nabla \varphi$.

Equations of motion obtained from the 
lagrangian (\ref{lagrangian_hydrodynamic}) have a standard hydrodynamic 
form
\begin{equation}
\label{kmuHydrSyst}
\begin{array}{c}
k_{T} \,\, = \,\, 8 \, J_{\perp} \, \mu \, \sin k \,\, k_{X}
\, - \, \left( 8 \, J_{\perp} \, \cos k \, - \, 8 \, J_{z}
\right) \, \mu_{X}  \cr
\mu_{T} \,\, = \,\, - \, 4 \, J_{\perp} \,
(1 - \mu^{2}) \, \cos k \,\, k_{X}
\, + \, 8 \, J_{\perp} \, \mu \, \sin k \,\, \mu_{X}
\end{array}
\end{equation}

\subsection{Linearized equations of motion. Stable and unstable regimes. }

Let us consider a  superfluid state with a uniform density and, possibly, 
finite phase winding. When $k \neq 0$, this is a current carrying state 
with $I = 4 J_\perp (1-\mu_0^2) \, \sin k$.

Frequencies of linearized excitations are given by the eigenvalues of 
the matrix
\begin{eqnarray}
A \,\, = \,\, \left(
\begin{array}{cc}
8 \, J_{\perp} \, \mu \, \sin k &
- \, 8 \, J_{\perp} \, \cos k \, + \, 8 \, J_{z} \cr
 - \, 4 \, J_{\perp} \, (1 - \mu^{2}) \, \cos k &
8 \, J_{\perp} \, \mu \, \sin k
\end{array}  \right)
\label{Amatrix}
\end{eqnarray}
We have for the eigenvalues of $A$
\begin{eqnarray}
\lambda_{1,2} \,\, = \,\, 8 \, J_{\perp} \, \mu \, \sin k 
\, \pm \, \sqrt{4 \, J_{\perp} \, (1 - \mu^{2}) \, \cos k \,
(8 \, J_{\perp} \, \cos k \, - \, 8 \, J_{z})}
\label{EigenValues}
\end{eqnarray}
When $J_{\perp} > J_{z}$ and $k$ is small, both eigenvalues of 
(\ref{Amatrix}) are real  (when $J_{z} = 0$ this is true for all $k$ ). 
This is the hyperbolic regime, which will be the main focus of our paper.

When $0 \, < \, \cos k \, < \, J_{z} / J_{\perp} $, 
eigenvalues of (\ref{Amatrix}) appear as a complex conjugate pair.
This is the elliptic regime, which corresponds to the unstable state of 
the system. In this regime small fluctuations of the plane wave type

$$\begin{array}{c}
k (X, T) \,\, = \,\, k_{0} \, + \, \delta k (X, T)
\,\,\,\,\,\,\,\, , \,\,\,\,\,\,\,\,
\mu (X, T) \,\, = \,\, \mu_{0} \, + \, \delta \mu (X, T) 
\cr
\delta k (X, T) \, \sim \, \delta k \, e^{i q X + i \nu (q) T}
\,\,\,\,\,\,\,\, , \,\,\,\,\,\,\,\,
\delta \mu (X, T) \, \sim \, \delta \mu \, 
e^{i q X + i \nu (q) T} 
\end{array} $$
grow exponentially in time. Existence of this unstable regime is known as 
the dynamical instability  \cite{Wu2001,Altman2005}. It was observed
experimentally for atoms in optical lattices\cite{Fallani2004,Mun2007}. 
Exponential growth of small modulations predicted by equations 
(\ref{kmuHydrSyst}) is  only valid for short times. Dynamics of
the unstable regime beyond the short time limit can be analyzed using 
mathematical methods from the theory of elliptic 
equations. In this paper we only address the stable hyperbolic regime.

When the initial state does not carry a current, i.e. there is no phase 
winding,
\begin{equation}
\label{phiInCond}
\varphi (X, 0) \,\, = \,\, 0
\end{equation}

To obtain further insight into the linearized system  we set
$$ \mu (X,T) \, = \, \mu_{0} + \rho (X,T) $$

We can now rewrite equation (\ref{kmuHydrSyst}) in terms of variables 
$\rho (X,T) $ and $k (X,T)$,
which describe small deviations from the equilibrium state


\begin{equation}
\label{LinkmuSyst}
k_{T} \,\, = \,\, - \, 8 \, (J_{\perp} - J_{z}) \, \rho_{X}
\,\,\,\,\,\,\,\, , \,\,\,\,\,\,\,\,
\rho_{T} \,\, = \,\, - \, 4 \, J_{\perp} \, (1 - \mu_{0}^{2}) 
\, k_{X}
\end{equation}

 System (\ref{LinkmuSyst}) gives the following equation for
$\rho (X, T)$

\begin{equation}
\label{LinrhoSyst}
\rho_{TT} \,\, = \,\, 32 \, J_{\perp} \, (1 - \mu_{0}^{2}) \,
(J_{\perp} - J_{z}) \, \rho_{XX}
\end{equation}
We find the familiar wave equation, which describes propagation
of the initial perturbation $\rho (X, 0)$ with a small amplitude.
Equations (\ref{LinkmuSyst}) and  (\ref{LinrhoSyst}) show that during the 
dynamical evolution of the perturbation,  the superfluid velocity 
$k (X, T)$ is of the order of $\rho (X, T)$, provided that this is true
in the initial state.  This is the regime that will be the focus
of our paper.


\subsection{Nonlinearities and appearance of singularities}

We now  include nonlinear terms in the analysis of  equations of motion. 
In the simplest case $J_{z} = 0$ we can define
\begin{eqnarray}
r^{1} \,\, = \,\, \sqrt{2} \, \arcsin \mu \, - \, k \,\, = \,\, 
\pi / \sqrt{2} \, - \, \sqrt{2} \, \theta \, - \, k
\nonumber\\
r^{2} \,\, = \,\, \sqrt{2} \, \arcsin \mu \, + \, k \,\, = \,\,
\pi / \sqrt{2} \, - \, \sqrt{2} \, \theta \, + \, k
\end{eqnarray}
And from the Lagrangian (\ref{lagrangian_hydrodynamic}) we obtain 
equations of motion
\begin{eqnarray}
\label{r1r2HydrSyst}
\begin{array}{c}
r^{1}_{T} \,\, = \,\, \left( 8 \, J_{\perp} \, 
\sin {r^{1} + r^{2} \over 2 \sqrt{2}} \, 
\sin {r^{2} - r^{1} \over 2} \, + \, 4 \sqrt{2} \, J_{\perp} \,
\cos {r^{1} + r^{2} \over 2 \sqrt{2}} \,
\cos {r^{2} - r^{1} \over 2} \right) \, r^{1}_{X} 
\cr
\cr
r^{2}_{T} \,\, = \,\, \left( 8 \, J_{\perp} \,
\sin {r^{1} + r^{2} \over 2 \sqrt{2}} \,
\sin {r^{2} - r^{1} \over 2} \, - \, 4 \sqrt{2} \, J_{\perp} \,
\cos {r^{1} + r^{2} \over 2 \sqrt{2}} \,
\cos {r^{2} - r^{1} \over 2} \right) \, r^{2}_{X}
\end{array}
\end{eqnarray}
Equations (\ref{r1r2HydrSyst}) are written in terms
of the Riemann invariants, which separate the system (\ref{kmuHydrSyst})
into the left- and  right-moving parts. This representation 
is most convenient in the analysis of Hydrodynamic Type systems.
System of equations (\ref{r1r2HydrSyst}) admits two natural
reductions $r^{1} = const$ or $r^{2} = const$, which describe 
separate propagation of the left- and right-moving excitations.

When $J_{z} \neq 0$,  expressions for Riemann
invariants are more cumbersome
$$r^{1,2} \,\, = \,\, \sqrt{2} \, \arcsin \mu \, - \,
\sqrt{2} \, \arcsin \mu_{0} \, \mp \, \int_{0}^{k}
{\sqrt{\cos k} \over \sqrt{\cos k - J_{z}/J_{\perp}}} \,\, d k $$
The corresponding diagonal system of the equations of motion
has a character close
to (\ref{r1r2HydrSyst}) in the hyperbolic regime.

 Taking in the account that the functions $\rho (X, T)$ and
$k (X, T)$ have the same order in our approach we can write

\begin{equation} 
\label{LambdaExp}
\begin{array}{c}
\lambda_{1,2} \,\, = \,\, \pm \, 
\sqrt{32 \, (1 - \mu_{0}^{2}) \, J_{\perp} \, (J_{\perp} - J_{z})} 
\,\, + \,\, 8 \, J_{\perp} \, \mu_{0} \, k \,\, \mp \,\, 
{\mu_{0} \over \sqrt{1 - \mu_{0}^{2}}} \,\,
\sqrt{32 \, J_{\perp} \, (J_{\perp} - J_{z})} \,\, \rho \,\, + 
\cr
\cr
+ \,\, 8 \, J_{\perp} \, \rho \, k \,\, \mp \,\,
\sqrt{2 \, (1 - \mu_{0}^{2}) \, J_{\perp}} \,\, 
{2 J_{\perp} - J_{z} \over \sqrt{J_{\perp} - J_{z}}} \,\, k^{2} 
\,\, \mp \,\, \sqrt{8 \, J_{\perp} \, (J_{\perp} - J_{z})} \,
{1 \over (\sqrt{1 - \mu_{0}^{2}})^{3}} \, \rho^{2} \,\, + \, \dots 
\end{array}
\end{equation}

 In the same way

\begin{equation}
\label{rhokExp}
r^{1,2} \, = \, {\sqrt{2} \over \sqrt{1 - \mu_{0}^{2}}} \,
\rho \, + \,
{\sqrt{2} \mu_{0} \over (\sqrt{1 - \mu_{0}^{2}})^{3}} \,
{\rho^{2} \over 2} \, + \,
{\sqrt{2} (1 + 2 \mu_{0}^{2}) \over (\sqrt{1 - \mu_{0}^{2}})^{5}}
\, {\rho^{3} \over 6} \, + \, \dots \, 
\mp \, {k \over \sqrt{1 - J_{z} / J_{\perp}}} \, \mp \,
{J_{z} \over 12 J_{\perp} ( \sqrt{1 - J_{z} / J_{\perp}} )^{3}}
\, k^{3} \, \mp \, \dots 
\end{equation}

 To understand the role of non-linear effects we expand 
the corresponding equations of motion up to  second order terms in 
deviations from the uniform state. We obtain the following  general form
of the equations of motion
 \begin{equation}
\label{rweaklynonlin}
\begin{array}{c}
r^{1}_{T} \,\, = \,\, \sqrt{J_{\perp} \, (J_{\perp} - J_{z})} \,
\left( \sqrt{ 32 \, (1 - \mu_{0}^{2})} \,\, - \,\,
6 \, \mu_{0} \, r^{1} \,\, + \,\, 2 \, \mu_{0} \, r^{2} \right)
\,\, r^{1}_{X}  \cr
r^{2}_{T} \,\, = \,\, \sqrt{J_{\perp} \, (J_{\perp} - J_{z})} \,
\left( - \, \sqrt{ 32 \, (1 - \mu_{0}^{2})} \,\, - \,\,
2 \, \mu_{0} \, r^{1} \,\, + \,\, 6 \, \mu_{0} \, r^{2} \right)
\,\, r^{2}_{X}
\end{array}
\end{equation}
Equations (\ref{rweaklynonlin}) describe coupled evolution of the
right and left moving parts. To get further insight into dynamics we 
make another simplification.
In the problems that we consider 
the left and right moving parts overlap at short times, but 
separate after a finite time. The main effects of non-linearities
appear at long times. Thus when discussing effects
of non-linearities it is sufficient to consider 
separately the left- and right-moving parts of the solution.
So when we discuss the dynamics
of $r^1$ we can set $r^2 = const$ and vice versa. 

After we make the Galilean transformation for the left and right 
propagating parts we obtain
\begin{equation}
\label{r1MovSyst}
r^{1}_{T} \,\, = \,\, - \,\, 6 \, \mu_{0} \,
\sqrt{J_{\perp} \, (J_{\perp} - J_{z})} \,\, r^{1} \,  r^{1}_{X}
\end{equation}
\begin{equation}
\label{r2MovSyst}
r^{2}_{T} \,\, = \,\, 6 \, \mu_{0} \,  
\sqrt{J_{\perp} \, (J_{\perp} - J_{z})} \,\, r^{2} \,  r^{2}_{X}
\end{equation}

\begin{figure}[t]
\begin{center}
\includegraphics[width=14.0cm,height=7cm]{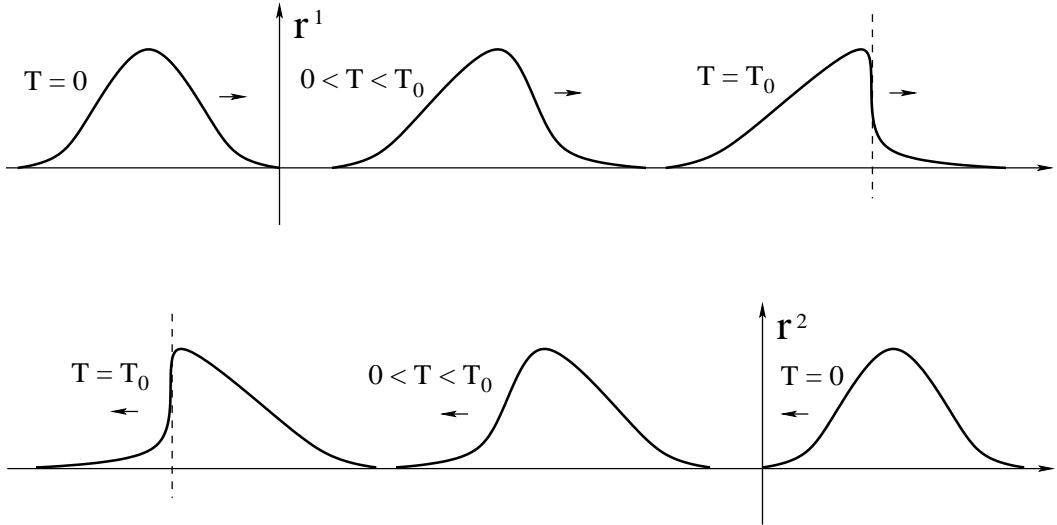}
\end{center}
\vspace{-.3in}
\caption{Formation of the breaking point for equations
(\ref{r1MovSyst}), (\ref{r2MovSyst}) in the case $\mu_{0} > 0$.
$r^1$ and $r^2$ describe left and right movers respectively.
Dynamics is shown in  moving frames of references.
In both cases  the rear edge of the wave steepens and develops a 
singularity. }
\label{rBreakPoint}
\end{figure}

Equations (\ref{r1MovSyst}) and (\ref{r2MovSyst})  
are known as the Hopf equations describing 
"simple waves". Their solutions 
are given by the implicit formula

$$r^{1,2} \,\, = \,\, F \left( X \, \mp \, 6 \, \mu_{0} \,
\sqrt{J_{\perp} \, (J_{\perp} - J_{z})} \,\, r^{1,2} \, T
\right) $$
The most important feature of these solutions  is that they 
exist only up to a finite time $T_0$, which depends on the
initial conditions. All nontrivial solutions become singular
after some finite time. Physically this corresponds to formation
of the breaking point, which we show in Fig.
\ref{rBreakPoint}. This can be understood
as a result of regions of different densities
moving with different velocities.

Formation of the singularity  is not
restricted to the truncated equations of motion (\ref{rweaklynonlin}).
This is a feature of the general non-linear dynamics
of the equations of motion (\ref{r1r2HydrSyst}).
Generally  system of equations (\ref{kmuHydrSyst})
can be reduced to a linear problem using the so-called hodograph 
transformation. Then solutions of (\ref{kmuHydrSyst})
can be described in terms of perturbations moving along the
characteristic lines $dX / dT = \lambda_{1,2} (r^{1}, r^{2})$.
Characteristics of the nonlinear system depend on the
variables $(r^{1}, r^{2})$ and unique solutions of
(\ref{kmuHydrSyst}) exist only up to a finite time $T_{0}$. At later 
times solution becomes multi-valued. Special solutions
of (\ref{kmuHydrSyst}) given by relations 
$r^{1} (X, T) = const$ or $r^{2} (X, T) = const$ describe 
perturbations moving along one of the characteristic lines. 
In this case the second variable ($r^{2}$ or $r^{1}$) satisfies a 
nonlinear first order equation, which is (locally) equivalent
to the nonlinear Hopf equation. 

For  times approaching $T_{0}$ solutions of 
(\ref{r1MovSyst}), (\ref{r2MovSyst}) are close to developing a breaking 
point and have  high gradients. In this regime neglecting higher order 
gradients in the Lagrangian (\ref{ApprLagr}) is no longer justified. 
In the next section we will see that  taking dispersion
into account suppresses singularities in the solutions and gives rise
to short-period oscillations.

General analysis of how dispersion leads to the formation of oscillatory 
zones in our system is rather
complicated. In the most generic case, one
can not use expansion (\ref{ApprLagr}) to describe the
oscillatory zone formation.
The period of oscillations arising
for $T > T_{0}$ is of the order of $h$, so all 
higher dispersive corrections are of the same order. Accurate
description of the transition from the "slowly-modulated" 
to the  rapidly modulated regimes
can only be done with the use of the original lattice
system (\ref{DiscreteSystem}).
However, there are certain special cases, in which the
use of the continuum model (\ref{ApprLagr}) is justified.
Fortunately these cases are  interesting from
the experimental point of view. They will be
the subject of our discussion.

\section{Nonlinear waves in generic case}

\subsection{Connection to Korteweg-de Vries equation}

When discussing dispersive terms for $\rho$ and $k$  in the equations of 
motion, it is sufficient to keep them only in the linear order in 
deviations from the uniform state. Dispersive terms come with additional 
factors of $h$ and are already small. Hence in the
Lagrangian (\ref{ApprLagr}) dispersive terms need to be considered
only up to quadratic terms in $\rho$ or 
$\sigma_{X}$. 
Modulo total derivatives with respect to $X$ we can write 

$${\cal L}_{Disp} \,\, \simeq \,\, h^{2} \, J_{\perp} \,
{\mu_{0}^{2} \over 1 - \mu_{0}^{2}} \,\, \rho_{X}^{2} \,\, + \,\,
h^{2} \, J_{z} \, \rho_{X}^{2} \,\, - \,\, {1 \over 12} \,
h^{2} \, J_{\perp} \, (1 - \mu_{0}^{2}) \,\, \sigma_{XX}^{2} $$

The resulting equations of motion are

\begin{equation}
\label{kmuDisp}
\begin{array}{c}
k_{T} \,\, \simeq \,\, 8 \, J_{\perp} \, \mu \, \sin k \,
k_{X} \, - \, \left( 8 \, J_{\perp} \, \cos k \, - \,
8 \, J_{z} \right) \, \mu_{X} \, + \, 4 \, h^{2} \, J_{\perp} \,
{\mu_{0}^{2} \over 1 - \mu_{0}^{2}} \,\, \rho_{XXX} \, + \,
4 \, h^{2} \, J_{z} \,\, \rho_{XXX} 
\cr
\cr
\mu_{T} \,\, \simeq \,\, - \, 4 \, J_{\perp} \,
\left( 1 - \mu^{2} \right) \, \cos k \,\, k_{X} \, + \,
8 \, J_{\perp} \, \mu \, \sin k \,\, \mu_{X} \, - \,
{1 \over 3} \,\, h^{2} \, J_{\perp} \, ( 1 - \mu_{0}^{2} ) \,\,
k_{XXX}
\end{array} 
\end{equation}

Using  variables $r^{1} (k, \mu)$, $r^{2} (k, \mu)$ we
obtain

\begin{equation}
\label{r1DispSyst}
\begin{array}{c}
r^{1}_{T} \,\, = \,\, \sqrt{J_{\perp} (J_{\perp} - J_{z})} \,
\left( \sqrt{32 ( 1 - \mu_{0}^{2} )} \, - \, 6 \, \mu_{0} \, r^{1}
\, + \, 2 \, \mu_{0} \, r^{2} \right) \, r^{1}_{X} \, - 
\cr
\cr
- \, {1 \over 3 \sqrt{2}} \, h^{2} \, 
\sqrt{J_{\perp} (J_{\perp} - J_{z})} \, \sqrt{1 - \mu_{0}^{2}} \,
\left( r^{2}_{XXX} \, - \, r^{1}_{XXX} \right) \, - 
\cr
\cr
- \, \sqrt{2} \, h^{2} \, \left( J_{\perp} \,
{\mu_{0}^{2} \over 1 - \mu_{0}^{2}} \, + \, J_{z} \right) \,
\sqrt{1 - \mu_{0}^{2}} \, 
\sqrt{{J_{\perp} \over J_{\perp} - J_{z}}} \, \left(
r^{1}_{XXX} \, + \, r^{2}_{XXX} \right) 
\end{array}
\end{equation}

\begin{equation}
\label{r2DispSyst}
\begin{array}{c}
r^{2}_{T} \,\, = \,\, \sqrt{J_{\perp} (J_{\perp} - J_{z})} \,
\left( - \, \sqrt{32 ( 1 - \mu_{0}^{2} )} \, - \, 2 \, \mu_{0} \, 
r^{1} \, + \, 6 \, \mu_{0} \, r^{2} \right) \, r^{2}_{X} \, - 
\cr
\cr   
- \, {1 \over 3 \sqrt{2}} \, h^{2} \,
\sqrt{J_{\perp} (J_{\perp} - J_{z})} \, \sqrt{1 - \mu_{0}^{2}} \,
\left( r^{2}_{XXX} \, - \, r^{1}_{XXX} \right) \, + 
\cr
\cr
+ \, \sqrt{2} \, h^{2} \, \left( J_{\perp} \,
{\mu_{0}^{2} \over 1 - \mu_{0}^{2}} \, + \, J_{z} \right) \,
\sqrt{1 - \mu_{0}^{2}} \,
\sqrt{{J_{\perp} \over J_{\perp} - J_{z}}} \, \left(
r^{1}_{XXX} \, + \, r^{2}_{XXX} \right) 
\end{array}
\end{equation}

As in our earlier discussion we consider separately the left and right 
moving parts, i.e. we take either $r^1= {\rm const}$ or $r^2= {\rm 
const}$. After we included effects of dispersion such reductions  are no 
longer exact. However, in cases of interest,  interaction between $r^{1}$ 
and $r^{2}$  gives rise only to small rapid oscillations. Such 
oscillations are expected to be much smaller than the structures that we 
discuss (see e.g. \cite{KrusZab}) and we will neglect them in this paper.
We also perform Galilean transformations for the two parts and obtain
in the moving coordinate systems

\begin{equation}
\label{r1disp}
r^{1}_{T} \,\, = \,\, - \, 6 \, \mu_{0} \,
\sqrt{J_{\perp} (J_{\perp} - J_{z})} \,\, r^{1} \,\, r^{1}_{X} 
\, + \, \sqrt{2} \, h^{2} \, \sqrt{1 - \mu_{0}^{2}} \,
\sqrt{{J_{\perp} \over J_{\perp} - J_{z}}} \,
\left( J_{\perp} \, \left( {1 \over 6} \, - \,
{\mu_{0}^{2} \over 1 - \mu_{0}^{2}} \right) \, - \,
{7 \over 6} \, J_{z} \right) \, r^{1}_{XXX} 
\end{equation}

\begin{equation}
\label{r2disp}
r^{2}_{T} \,\, = \,\, 6 \, \mu_{0} \,
\sqrt{J_{\perp} (J_{\perp} - J_{z})} \,\, r^{2} \,\, r^{2}_{X} 
\, - \, \sqrt{2} \, h^{2} \, \sqrt{1 - \mu_{0}^{2}} \,
\sqrt{{J_{\perp} \over J_{\perp} - J_{z}}} \,
\left( J_{\perp} \, \left( {1 \over 6} \, - \,
{\mu_{0}^{2} \over 1 - \mu_{0}^{2}} \right) \, - \,
{7 \over 6} \, J_{z} \right) \, r^{2}_{XXX} 
\end{equation}

Note that equations (\ref{r1disp}), (\ref{r2disp}) transform into each
other if we change $X \rightarrow - X$. Equivalence of the  two equations 
for  fixed values of $J_{\perp}$, $J_{z}$ and $\mu_{0}$
represents an evident corollary of the symmetry $X \rightarrow - X$
of the original system.

 It is not difficult to see that equations (\ref{r1disp}), (\ref{r2disp}) 
represent the KdV-equation provided that
$$\mu_{0} \, \neq \, 0 \,\,\,\,\,\,\,\, , \,\,\,\,\,\,\,\,
J_{\perp} \, \left( {1 \over 6} \, - \,
{\mu_{0}^{2} \over 1 - \mu_{0}^{2}} \right) \, - \,
{7 \over 6} \, J_{z} \, \neq \, 0 $$

Depending on the values of parameters $J_{\perp}$, $J_{z}$, and $\mu_{0}$,
equations  (\ref{r1disp}) and (\ref{r2disp})
are equivalent to one of the following two equations

\begin{equation}
\label{minus}
U_{T} \, + \, 6 \, U \, U_{X} \, - \, U_{XXX} \, = \, 0
\end{equation}
\begin{equation}
\label{plus}
U_{T} \, + \, 6 \, U \, U_{X} \, + \, U_{XXX} \, = \, 0
\end{equation}
after an appropriate rescaling of coordinates $(X, T)$ and 
functions $r^{i}$. These two equations  are equivalent to each other
if we admit the inversion $r^{i} \rightarrow - r^{i}$,
$T \rightarrow - T$. However,
this transformation  leads to very different physical interpretation
of solutions for a fixed $\mu_{0}$, as we discuss
below.\footnote{In the next chapter we will also discuss 
that solutions of (\ref{minus}) and (\ref{plus}) demonstrate 
different stability properties
with respect to two-dimensional modulations.}

\begin{figure}
\begin{center}
\includegraphics[width=14.0cm,height=9cm]{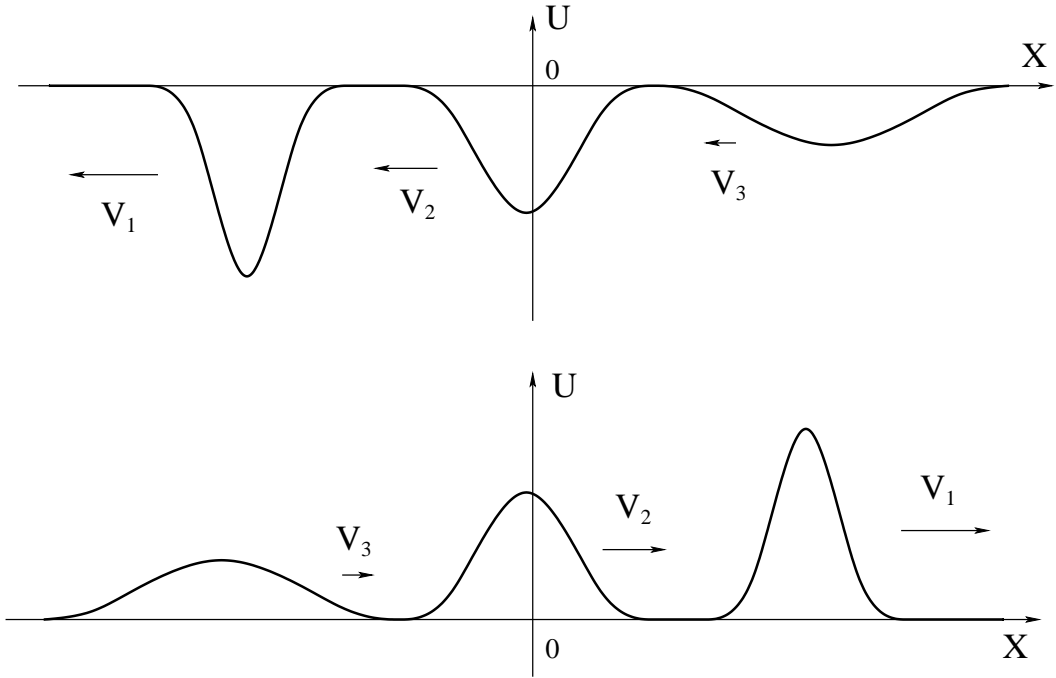}
\end{center}
\caption{The asymptotic form ($T \rightarrow \infty$) of the
$N$-soliton solutions for equations (\ref{minus}) and
(\ref{plus}) respectively ($V_{1} > V_{2} > V_{3}$).}
\label{Nsoliton}
\end{figure}

\begin{figure}
\begin{center}
\includegraphics[width=14.0cm,height=9cm]{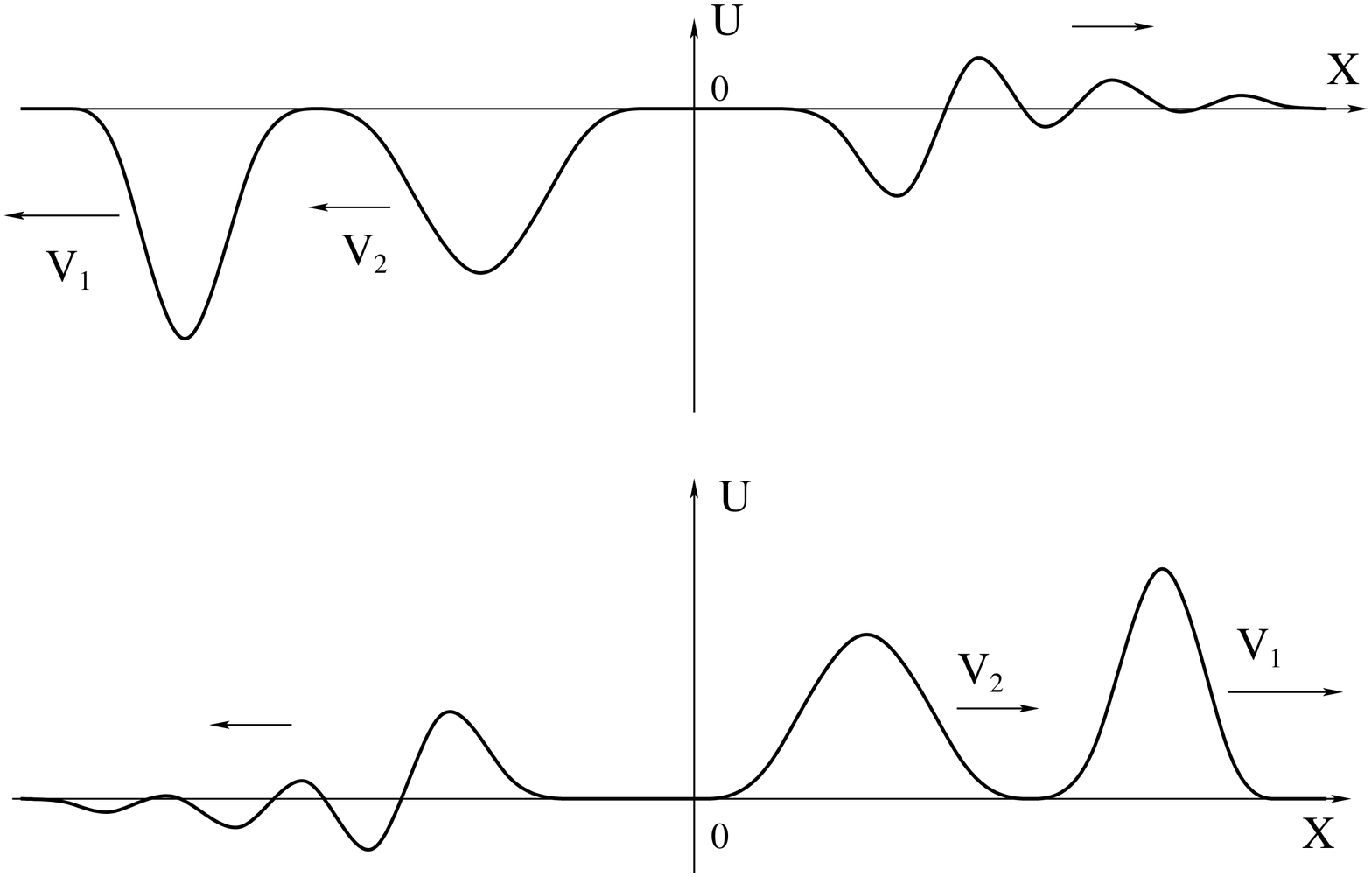}
\end{center}
\caption{The asymptotic form ($T \rightarrow \infty$) of the
general solution for equations (\ref{minus}) and (\ref{plus}) 
respectively.}
\label{GenSol}
\end{figure}

KdV type  equations (\ref{minus}) and (\ref{plus}) allow 
solitonic solutions, which are long lived
nonlinear excitations in the system.
The velocity of a soliton is proportional to its amplitude,
so larger solitons  move faster than the smaller ones.
The asymptotic form of an $N$-soliton solution for
$T \rightarrow \infty$ for equations (\ref{minus}) and
(\ref{plus}) can be represented as shown at Fig. \ref{Nsoliton}.
In the Appendix \ref{appendix_kdv}
we briefly review how one can
verify the existence of solitonic excitations in the KdV equation
using connection to the linear Schroedinger equation.
We also point out that in general,  solutions of (\ref{minus}) and
(\ref{plus}) include not only the soliton part
but also "wave trains". The soliton part and the
"wave train" parts
separate from each other at long times
(Fig. \ref{GenSol}). The soliton part of the solution remains unchanged 
for all $T > 0$ while the wave - train part "dissolves" as
$T \rightarrow \infty$ (\cite{ZakhMan}). From our point of view, 
solitons of
(\ref{minus}) and (\ref{plus}) represent the most interesting 
part of the solution and we focus on them in this paper.

\subsection{Discussion of solitonic excitations}

 Solitons in KdV equations have been studied in detail during the last
few decades. In this paper we take  previously known mathematical 
results and discuss their physical implications for our specific system. 
While we  provide a brief summary of the mathematical methods used in 
analyzing soliton excitations  in the Appendix,
we refer readers to the books \cite{AblSeg,Newell,NovManPitZakh} for
a more detailed discussion of  general mathematical aspects
of the KdV equation.

 The character of solitonic solutions of KdV type  equations 
(\ref{minus}) and (\ref{plus}) depends on  parameters. 
In particular
depending on the ratio of $J_z/J_\perp$ and the density, 
isolated solitons can appear either as particle-like or hole-like 
excitations. In this subsection we only provide a summary of the 
results. More details can be found in the Appendix.

\begin{figure}
\begin{center}
\includegraphics[width=14.0cm,height=7cm]{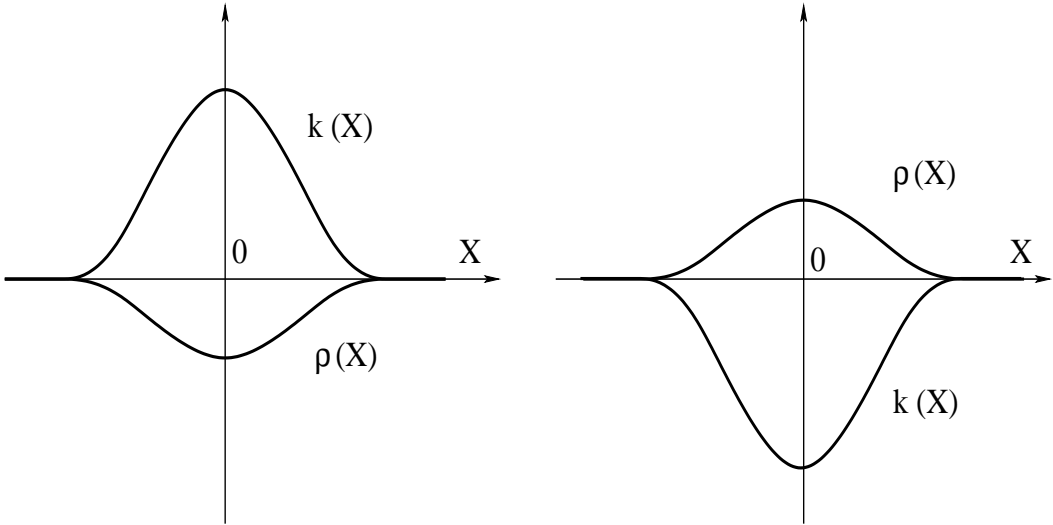}
\end{center}
\caption{The form of a soliton solution in the
$(k, \rho)$-variables for the case of the "hole-like"
soliton and the "particle-like" soliton respectively.}
\label{kmuSol}
\end{figure}

In the discussion below we only consider the case $\mu_0>0$, which
corresponds to the density above half-filling $\langle n \rangle > 1/2$.
Equations (\ref{r1disp}) and  (\ref{r2disp}) have a symmetry 
$\mu_0 \rightarrow - \mu_0$, $r_i \rightarrow - r_i$. This symmetry 
originates from the particle-hole symmetry of the initial system,
which relates states below and above half-filling
(\ref{DiscreteSystem}) :
$\theta \, \rightarrow \, \pi \, - \, \theta$ and
$\varphi \, \rightarrow \, - \, \varphi $.
In our discussion this symmetry allows to relate solitonic excitations 
below and above half-filling. For $\mu_0 < 0$ solitons are "mirror images"  
of the $\mu_0>0$ case. For example, if we find particle-like solitons 
above half-filling, we should have hole-like solitons below 
half-filling ($\mu_0 \rightarrow - \mu_0$) for the same values of $J$.
Let us represent here also the form of the "hole-like" and the 
"particle-like" solitons in the original variables $(k, \rho)$
(see Fig. \ref{kmuSol}).

We also remind the readers that we only need to consider states that
are stable against dynamical modulations, i.e. $J_{\perp} > J_{z}$.

\subsubsection{Solitons for  $ J_{z} > J_{\perp}/7$ and $\mu_0 > 0$. }

In this case both  equations
(\ref{r1disp}), (\ref{r2disp}) reduce to equation (\ref{plus}) after
rescaling the variables and, if necessary, performing the transformation 
$X \rightarrow - X$.
There should be no solitons  when $U (X) \leq 0$.
We are guaranteed to find solitonic excitations when 
\begin{equation}
\label{PlusInt}
\int_{-\infty}^{+\infty} U (X) \,\, d X \,\, > \,\, 0
\end{equation}
When $U (X) \geq 0$ the soliton part represents the main part of the 
solution. So we find particle-like solitons in this situation.

\subsubsection{
Solitons for 
$ J_{z} < J_{\perp}/7$ and $0 \,\, < \,\, \mu_{0} \,\, < \,\,   
\sqrt{{J_{\perp} - 7 J_{z} \over 7 (J_{\perp} - J_{z})}} $}

Now both equations (\ref{r1disp}), (\ref{r2disp}) 
reduce to equation (\ref{minus}) after rescaling the variables and 
doing the transformation $X \rightarrow - X$ in equation (\ref{r2disp}). 
This equation does not have any solitons when $U (X) \geq 0$. It has 
guaranteed solitonic solutions when 
\begin{equation}
\label{MinusInt}
\int_{-\infty}^{+\infty} U (X) \,\, d X \,\, < \,\, 0
\end{equation}
In the case with $U (X) \leq 0$ the soliton part represents the main part 
of the solution. Hence in terms of the original density, we find the 
hole-type solitons in this case.

\subsubsection{Solitons for $ J_{z} < J_{\perp}/7$ and 
$\sqrt{{J_{\perp} - 7 J_{z} \over 7 (J_{\perp} - J_{z})}}
\,\, < \,\, \mu_{0} \,\, < \,\, 1$}

 Both the equations (\ref{r1disp}), (\ref{r2disp}) reduce to equation 
(\ref{plus}) in this case. Thus we find particle-like solitons in terms of 
the original density.

We can now summarize results of this subsection.
When $ J_{z} > J_{\perp}/7$ we find that above half-filling there are only 
particle-like solitonic excitations. When $ J_{z} < J_{\perp}/7$ and above 
half-filling  we find that we have either hole-like (closer to half-filling)
or particle-like solitons (closer to filling factor one).

\subsection{Self-consistency of the long wavelength  expansion}

Before concluding this section we would like to verify that
our solutions do not take us outside the region of
applicability of 
Lagrangian (\ref{ApprLagr}), which was obtained using long wavelength 
expansion. When we consider dynamics starting from a state with 
small smooth deviations from a uniform density, approximate 
Lagrangian (\ref{ApprLagr}) can be certainly used at the initial 
stages of the evolution. However, at final
(asymptotic) stages of the evolution, the 
solution may be sufficiently different from the
initial state. Let us consider specifically
the soliton part of  asymptotic solutions. 
In soliton solutions both the nonlinear and dispersive
parts are important and the interplay of the two gives rise 
to a stable soliton. 
One of the important properties of the KdV equation 
is that the amplitude of solitons is of the same order as 
initial deviations from the uniform density.
Equations  (\ref{r1disp}) - (\ref{r2disp}) were obtained
assuming small deviations of the initial density 
from the uniform value $\mu_{0}$. These small deviations 
set the scale for the amplitude
of resulting solitons. In solitons there is a direct relation between the 
amplitude and the width (the width increases  as the amplitude goes to 
zero). Hence in the limit that we discuss, the dispersion part of our 
soliton solutions should  be small, and our approximation of neglecting 
higher dispersion corrections should be justified even for 
the final stages of the evolution. For example, when solution can be 
written as the "quasiclassical solution", in which the soliton 
part represents the main contribution to the solution,  higher 
dispersive and nonlinear terms should have very weak effect 
on the soliton.

Similar considerations are applicable  for the
"wave-train" part of the solutions. However, the
"wave-train" part dissolves  in the limit
$T \rightarrow \infty$ and we expect that it will be more
challenging to observe it in experiments.

\section{Nonlinear waves in special cases}

\subsection{Half-filling. Solitons of the modified  Korteweg-de Vries 
equation}

When the particle density is $1/2$, the system of hard core bosons
has a full particle-hole symmetry.
Eigenvalues $\lambda_{1,2}$ of the linearized system
(\ref{LinkmuSyst}) have  the largest possible magnitude

$$\lambda_{1,2} \,\, = \,\, \pm \, 4 \, 
\sqrt{2 J_{\perp} ( J_{\perp} - J_{z} )} $$ 
which corresponds to the largest possible velocity of linear waves. 
In the case of dynamics starting from some initial state, this should 
provide fastest spatial separation of the left- and 
right-moving parts of the perturbation. In this case $\mu_{0} = 0$,
so corrections to $\lambda_{1,2}$, which are linear  in $\rho$ and $k$, 
vanish and we need to use quadratic terms in the expansion 
(\ref{LambdaExp}). In our discussion below we keep linear terms, in order
to accommodate  small $\mu_{0} \neq 0$. Using approximation 
(\ref{rhokExp}) we can write

$$\lambda_{1} \simeq \sqrt{J_{\perp} (J_{\perp} - J_{z})}
\left[ 4 \sqrt{2} - 6 \mu_{0} r^{1} + 2 \mu_{0} r^{2} -  
{7 J_{\perp} - J_{z} \over 2\sqrt{2} J_{\perp}} (r^{1})^{2}
 +  {J_{\perp} + J_{z} \over 2\sqrt{2} J_{\perp}}
(r^{2})^{2} + {J_{\perp} - J_{z} \over 2\sqrt{2} J_{\perp}}
r^{1} r^{2} \right] $$
$$\lambda_{1} \simeq \sqrt{J_{\perp} (J_{\perp} - J_{z})}
\left[ - 4 \sqrt{2} - 2 \mu_{0} r^{1} + 6 \mu_{0} r^{2} -
{J_{\perp} + J_{z} \over 2\sqrt{2} J_{\perp}} (r^{1})^{2}
 +  {7 J_{\perp} - J_{z} \over 2\sqrt{2} J_{\perp}}
(r^{2})^{2} - {J_{\perp} - J_{z} \over 2\sqrt{2} J_{\perp}}
r^{1} r^{2} \right] $$

From the last two equations we determine how propagation of the left- 
and right- moving parts, (\ref{r1disp})-(\ref{r2disp}), is modified 
by the higher order terms. Within the assumptions of spatial 
separation of the left- and right-moving parts, 
which we used in the earlier discussion, and using appropriate moving
frames of reference we find

\begin{equation}
\label{r1next}
r^{1}_{T} \, = \, \sqrt{J_{\perp} ( J_{\perp} - J_{z} )} 
\, \left( - 6 \mu_{0} r^{1} - 
{7 J_{\perp} - J_{z} \over 2\sqrt{2} J_{\perp}}
(r^{1})^{2} \right) \, r^{1}_{X} \, + \, h^{2} \,
\sqrt{{2 J_{\perp} \over 6 (J_{\perp} - J_{z})}}
\, ( J_{\perp} - 7 J_{z} ) \, r^{1}_{XXX}
\end{equation}
\begin{equation}
\label{r2next}
r^{2}_{T} \, = \, \sqrt{J_{\perp} ( J_{\perp} - J_{z} )}
\, \left( 6 \mu_{0} r^{2} +
{7 J_{\perp} - J_{z} \over 2\sqrt{2} J_{\perp}}
(r^{2})^{2} \right) \, r^{2}_{X} \, - \, h^{2} \,
\sqrt{{2 J_{\perp} \over 6 (J_{\perp} - J_{z})}}
\, ( J_{\perp} - 7 J_{z} ) \, r^{2}_{XXX}
\end{equation}
In writing the last equations we omitted higher order corrections in 
$\mu_{0}$. 
When $J_{\perp} \neq 7 J_{z}$. equation (\ref{r1next}) can be
written in the canonical form
\begin{equation}
\label{mKdV}
U_{T} \, + \, \left( \alpha \, U \, + \, 6 \, U^{2} \right) \,
U_{X} \, \pm \, U_{XXX} \,\, = \,\, 0
\end{equation}
after a scaling transformation. Parameter $\alpha$ that
we introduced here is proportional to the deviation
from half-filling, $\alpha \sim \mu_{0}$, and we  assume it to be
small.
 
 Equation (\ref{mKdV}) is called the modified Korteweg - de Vries
(mKdV) equation and represents an integrable system as well as
the KdV equation (see \cite{Wadati}). Let us note also that the
mKdV equation is connected with the KdV equation by the Miura
transformation (\cite{Miura}) which was the first observation
of the integrability properties of the KdV equation itself
(see \cite{Newell}).

There is a wider variety of soliton excitations that one can construct 
in the mKdV problem. At a fixed value of the chemical potential 
one can find both particle-like and hole-like solitons moving in the same 
direction. This should be contrasted to the situation away from 
half-filling, which we discussed in the previous section, where at a 
given chemical potential and direction of propagation we had either 
particle or hole like solitons, but never both simultaneously. For the 
mKdV case we also find soliton excitations  which look like
particle on a pedestal (or hole on a pedestal). We provide a detailed 
discussion of solitons in the mKdV problem and their manifestations for 
our system in the Appendix.

\subsection{Close to integer filling. Nonlinear Schroedinger equation}

When the system is close to integer filling
$\mu_{0} = \pm 1$.
In this case characteristic velocities of the linearized system
(\ref{LinkmuSyst}) coincide. Hence we can no longer assume separation 
of the left- and right-moving parts. Examining system 
(\ref{r1DispSyst})-(\ref{r2DispSyst}) we find
that dispersive corrections also have singularities in variables
$(\rho , k)$ .

To avoid these difficulties we return to 
variables $(\theta , \sigma)$, which we used before, 
and consider the Lagrangian density 

\begin{equation}
\label{ThetaSigmaLagr}
\begin{array}{c}
{\cal L} \,\, = \,\, {1 \over 2} \, \sigma_{T} \, \cos \theta
\, - \, 2 \, J_{\perp} \, \sin^{2} \theta \, \cos \sigma_{X}
\, - \, 2 \, J_{z} \, \cos^{2} \theta \, - \, h^{2} \,
J_{\perp} \, \sin \theta \, \left( \sin \theta \right)_{XX}
\, \cos \sigma_{X} \, - 
\cr
\cr
- \, h^{2} \, J_{\perp} \, \sin^{2} \theta \, \left(
{1 \over 6} \, \sigma_{XXX} \, \sin \sigma_{X} \, + \,
{1 \over 4} \, \sigma_{XX}^{2} \, \cos \sigma_{X} \right) 
\, - \, h^{2} \, J_{z} \, \cos \theta \,  \left(
\cos \theta \right)_{XX} \,\, + \,\, {\cal O} (h^{4}) 
\end{array}
\end{equation}
in the limit $\theta \rightarrow 0, \, \pi$. If we keep only 
quadratic terms in $(\sin \theta , \sigma)$ in the dispersive
part of the Lagrangian, we can write the corresponding equations
of motion as (in the limit $\mu_{0} = \pm 1$)

\begin{equation}
\label{ThetaSigmaSyst}
\begin{array}{c}
\sigma_{T} \, \sin \, \theta \, + \, 8 \, J_{\perp} \,
\sin \, \theta \, \cos \, \theta \, \cos \, \sigma_{X} \, - \,
8 \, J_{z} \, \cos \, \theta \, \sin \, \theta \, + \,
4 \, h^{2} \, J_{\perp} \, \mu_{0} \, \left( 
\sin \, \theta \right)_{XX} \,\, = \,\, 0
\cr
\cr
\theta_{T} \, - \, 8 \, J_{\perp} \, \left(
\sin \, \theta \right)_{X} \, \sin \, \sigma_{X} \, - \,
4 \, J_{\perp} \, \sin \, \theta \, \left( 
\sin \, \sigma_{X} \right)_{X} \,\, = \,\, 0
\end{array}
\end{equation}

If we keep only the lowest order cubic terms in the nonlinear part of 
(\ref{ThetaSigmaSyst}), we can rewrite this equation for small 
$(\theta , \sigma)$ as

$$ \begin{array}{c}
\sigma_{T} \, \sin \, \theta \, + \, 8 \, \mu_{0} \, \left( 
J_{\perp} - J_{z} \right) \, \sin \, \theta \, - \,
4 \, \mu_{0} \, \left( J_{\perp} - J_{z} \right) \,
\sin^{3} \, \theta \, - \, 4 \, \mu_{0} \, J_{\perp} \,
\sin \, \theta \, \sigma_{X}^{2} 
\, + \, 4 \, h^{2} \, \mu_{0} \, J_{\perp} \, \left(
\sin \, \theta \right)_{XX} \,\, = \,\, 0 
\cr
\cr
\theta_{T} \, - \, 8 \, J_{\perp} \, \left(
\sin \, \theta \right)_{X} \, \sigma_{X} \, - \,
4 \, J_{\perp} \, \sin \, \theta \,\,
\sigma_{XX} \,\, = \,\, 0 
\end{array} $$
($\mu_{0} = \pm 1$).

 It is not difficult to verify that the system above can be 
written in the form of the defocusing nonlinear Shr\"odinger
equation

\begin{equation}
\label{NonLinShr}
i h \, \psi_{T} \,\, = \,\, 8 \, \left( J_{\perp} - J_{z} \right) 
\, \psi \, - \, 4 \, \left( J_{\perp} - J_{z} \right) \,
|\psi|^{2} \psi \, + \, 4 \, h^{2} \, J_{\perp} \, \psi_{XX}
\end{equation}
for the function

$$\psi \,\, = \,\, \sin \, \theta \,\, 
e^{i \mu_{0} \sigma / h} \,\, = \,\,
\sin \, \theta \,\, e^{i \mu_{0} \varphi} $$

 Equation (\ref{NonLinShr}) describes
an integrable system (\cite{ZakhShab}), which was solved by
V.E. Zakharov and A.B. Shabat by the inverse scattering method.
System (\ref{NonLinShr}) admits an exact description of the
evolution starting from any initial state. However, nonlinear 
Shr\"odinger equation does not have soliton solutions in the defocusing 
case. Defocusing nature of equation (\ref{NonLinShr}) demonstrates 
stable behavior of the system with respect to initial perturbations. 
In this case asymptotic behavior of solutions of (\ref{NonLinShr}) 
should only include wave-trains which "dissolve" for $T \rightarrow 
\infty$ (\cite{ZakhMan}). \footnote{The soliton solutions on the
"pedestal" are also possible for equation (\ref{NonLinShr}).
We do not consider them here.}

\subsection{Special filling factor}

We now comment on
the special point of our system at

\begin{eqnarray}
\mu_{0} \,\, = \,\, \sqrt{J_{\perp} - 7 J_{z}} \, / \,
\sqrt{7 (J_{\perp} - J_{z})} 
\label{muSpecial}
\end{eqnarray}
for the case $0 < J_{z} < J_{\perp}/7$. To get
equations (\ref{minus}) - (\ref{plus}) from 
(\ref{r1disp}) - (\ref{r2disp}) we need to make scaling
transformation

$$X \,\, \rightarrow \,\, X \, \left/ \, h \, \left(
{2 (1 - \mu_{0}^{2}) J_{\perp} \over J_{\perp} - J_{z}}
\right)^{1/4} \, \left| J_{\perp} \, \left( {1 \over 6} \, - \,
{\mu_{0}^{2} \over 1 - \mu_{0}^{2}} \right) \, - \,
{7 \over 6} \, J_{z} \right|^{1/2} \right. $$
This transformation is singular at the special point (\ref{muSpecial}). 
As a corollary, the width
of solitons (and the period of oscillations in the "wave-train"
part) become small in $X$-space w.r.t. another parameter

$$\mu_{0} \,\, - \,\, \sqrt{J_{\perp} - 7 J_{z}} \, / \,
\sqrt{7 (J_{\perp} - J_{z})} $$

Higher dispersive terms
become important in this limit and Lagrangian density (\ref{ApprLagr}) 
can no longer be used. As we discussed earlier, dynamics is more 
complicated near this special point, and one should use original 
lattice system (\ref{DiscreteSystem}) to discuss dynamics. In general
we expect here oscillation zones with rather short period of
oscillations.

\section{Decay of the density step}

\begin{figure}
\begin{center}
\includegraphics[width=14.0cm,height=5cm]{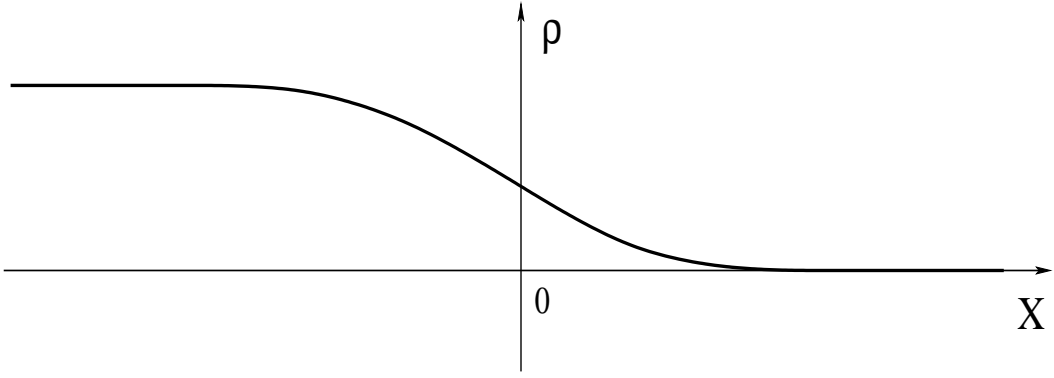}
\end{center}
\caption{The step-like initial conditions for the function
$\rho (X, T)$ with the assumption $\varphi (X, 0) = 0$.}
\label{StepInCond}
\end{figure}

We now apply our general arguments to understand dynamical evolution 
starting from a specific initial state. We assume that at $T=0$
we have a smooth step-like change in the density without any initial
current. Experimentally such initial configuration can be created using a 
smooth step in the external potential that is suddenly removed. This 
initial state is of the form given by equation  (\ref{phiInCond}).
It is shown schematically in Fig. \ref{StepInCond}. In this section we 
only consider the situation when the system is not close
to any special points. Density step decay for systems close to 
half-filling is discussed in the Appendix.

\begin{figure}
\begin{center}
\includegraphics[width=14.0cm,height=6cm]{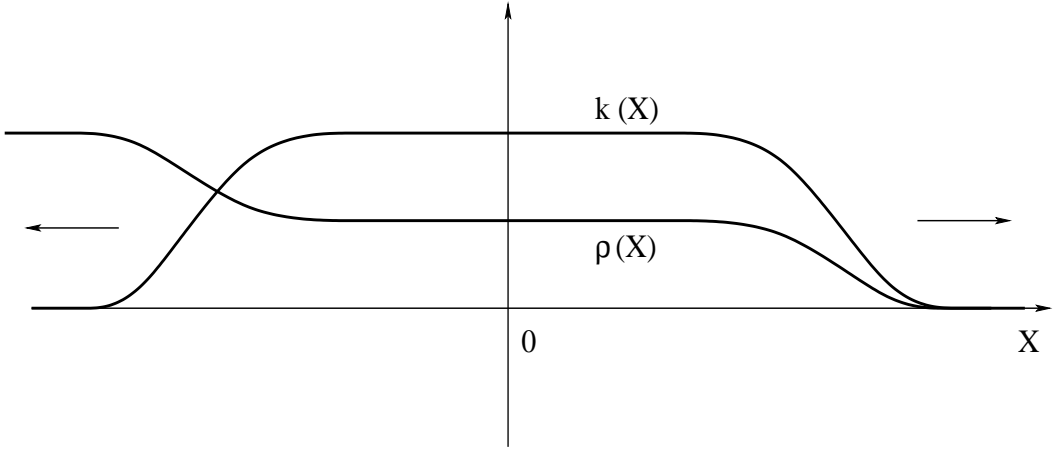}
\end{center}
\caption{The evolution of initial distribution with  
$k (X, 0) = 0$ and the step-like $\rho (X, 0)$ in the
approximation of system (\ref{LinkmuSyst}).}
\label{TwoSteps}
\end{figure}

Since we rely on the long wavelength expansion, we 
assume that function $\rho \, (X, T=0)$ is a slow function of 
the spatial coordinate.

The main terms in the long wavelength expansion of dynamics
are given by the wave equation (\ref{LinkmuSyst}). The wave equation 
predicts that  after a short  time
the step-like initial state should turn into a two-step solution,
with two steps propagating in the opposite directions (see Fig. 
\ref{TwoSteps}). When the two steps separate from each other, they can 
be analyzed independently. The left- and  
right-moving edges of the solution correspond to $r_1$ and $r_2$. 
Proceeding to the next order in $h$, we find that
they are described by equations  (\ref{r1disp}) and (\ref{r2disp}) 
respectively. After rescaling of
coordinates $(X, T)$ and functions $r^{1}$ and $r^{2}$ themselves,
this dynamics is given 
either by equation (\ref{minus}) or (\ref{plus}), where the choice 
depends on the values  of $(J_{\perp}$, $J_{z}$,  and $\mu_{0})$.

\begin{figure}
\begin{center}
\includegraphics[width=14.0cm,height=8cm]{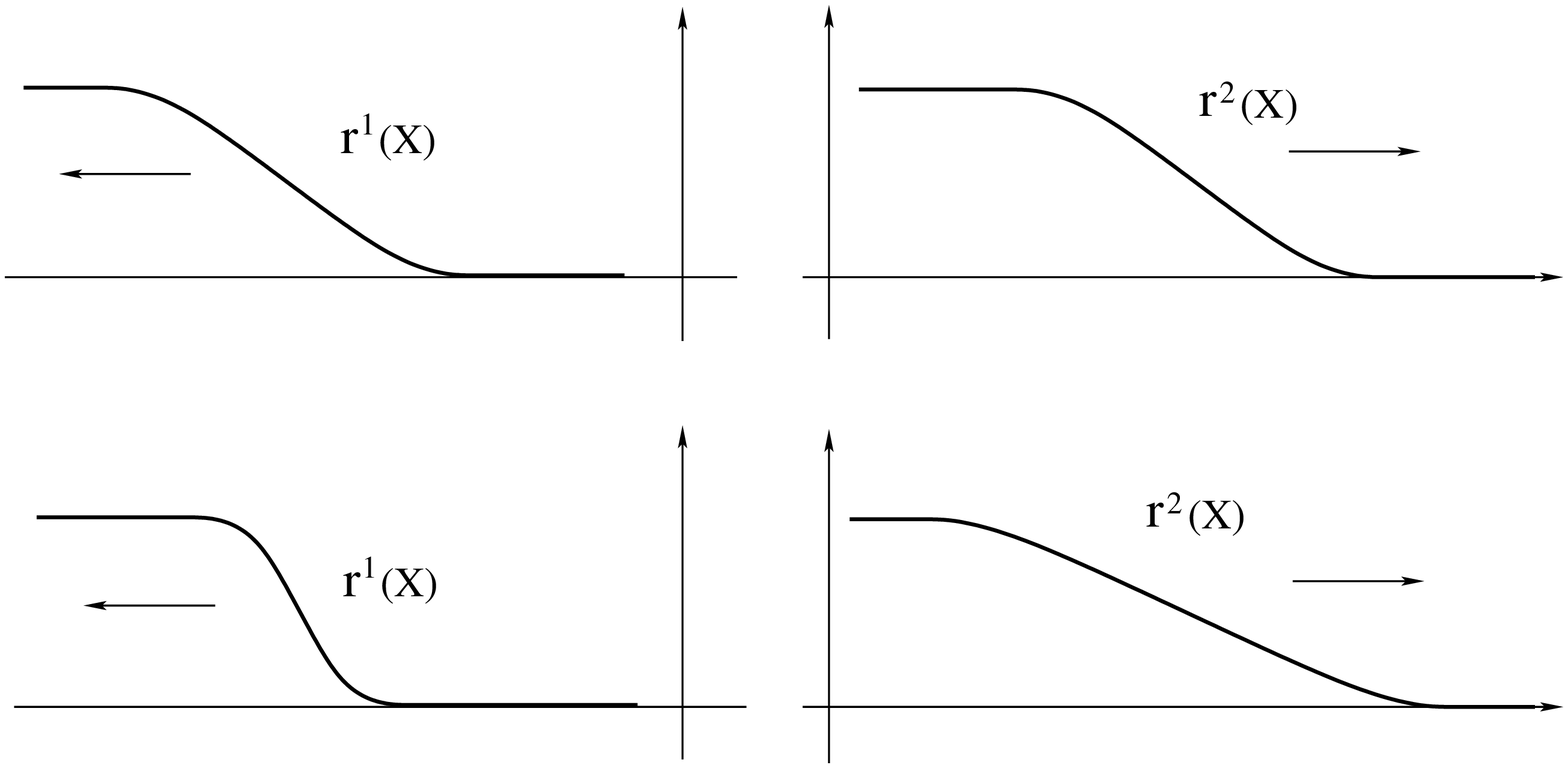}
\end{center}
\caption{The increasing of the steepness of solution $r^{1} (X)$
on the left-moving edge and the decreasing of the steepness of
solution $r^{2} (X)$ on the right-moving edge in the
hydrodynamic approximation. The data are sketched in the
original coordinate system.}
\label{r1r2Hydr}
\end{figure}

\begin{figure}
\begin{center}
\includegraphics[width=14.0cm,height=8cm]{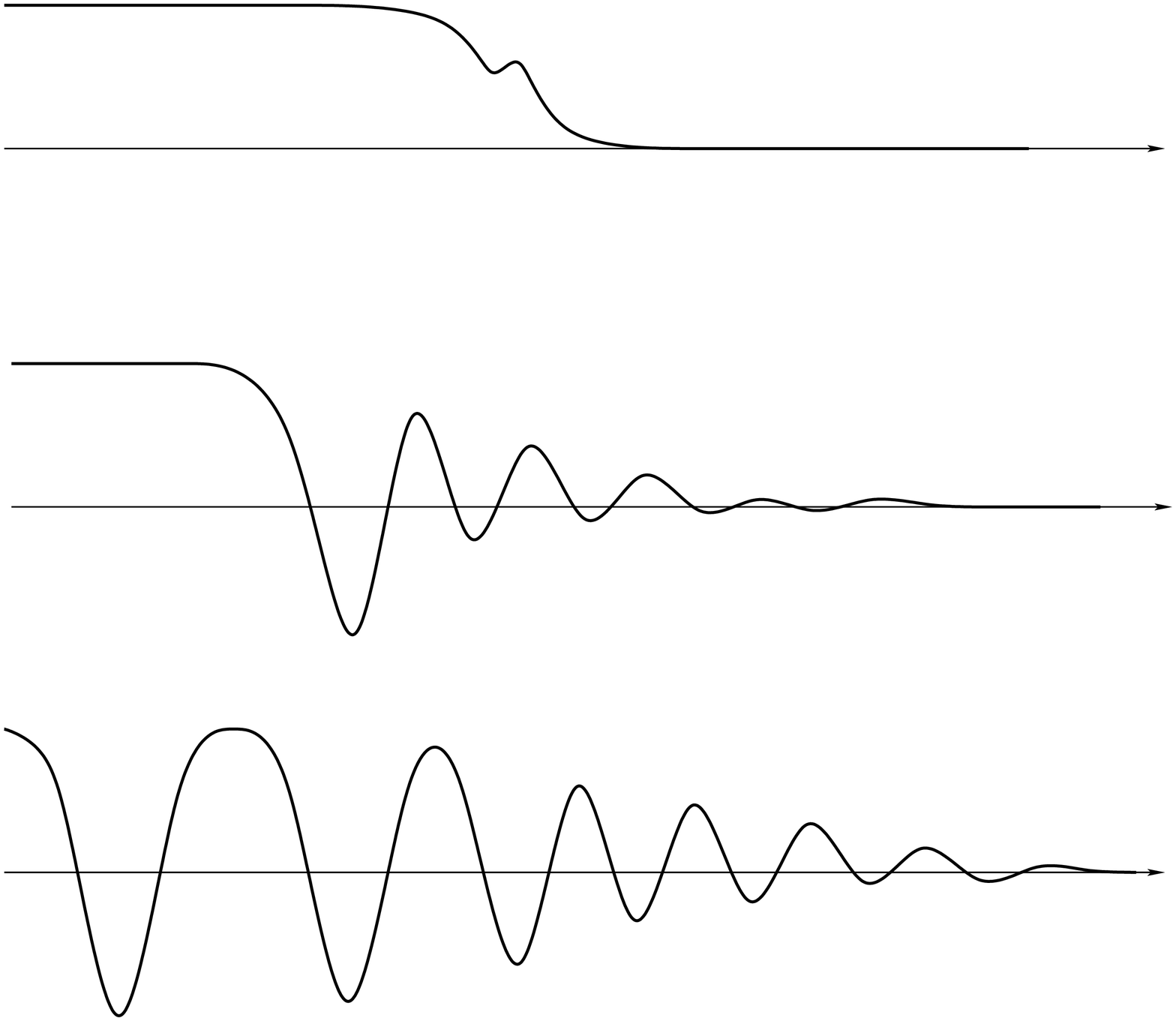}
\end{center}
\caption{The development of oscillation zone from the step-like
initial data for the case of equation (\ref{minus}).}
\label{StepOscMinus}
\end{figure}

\begin{figure}
\begin{center}
\includegraphics[width=14.0cm,height=8cm]{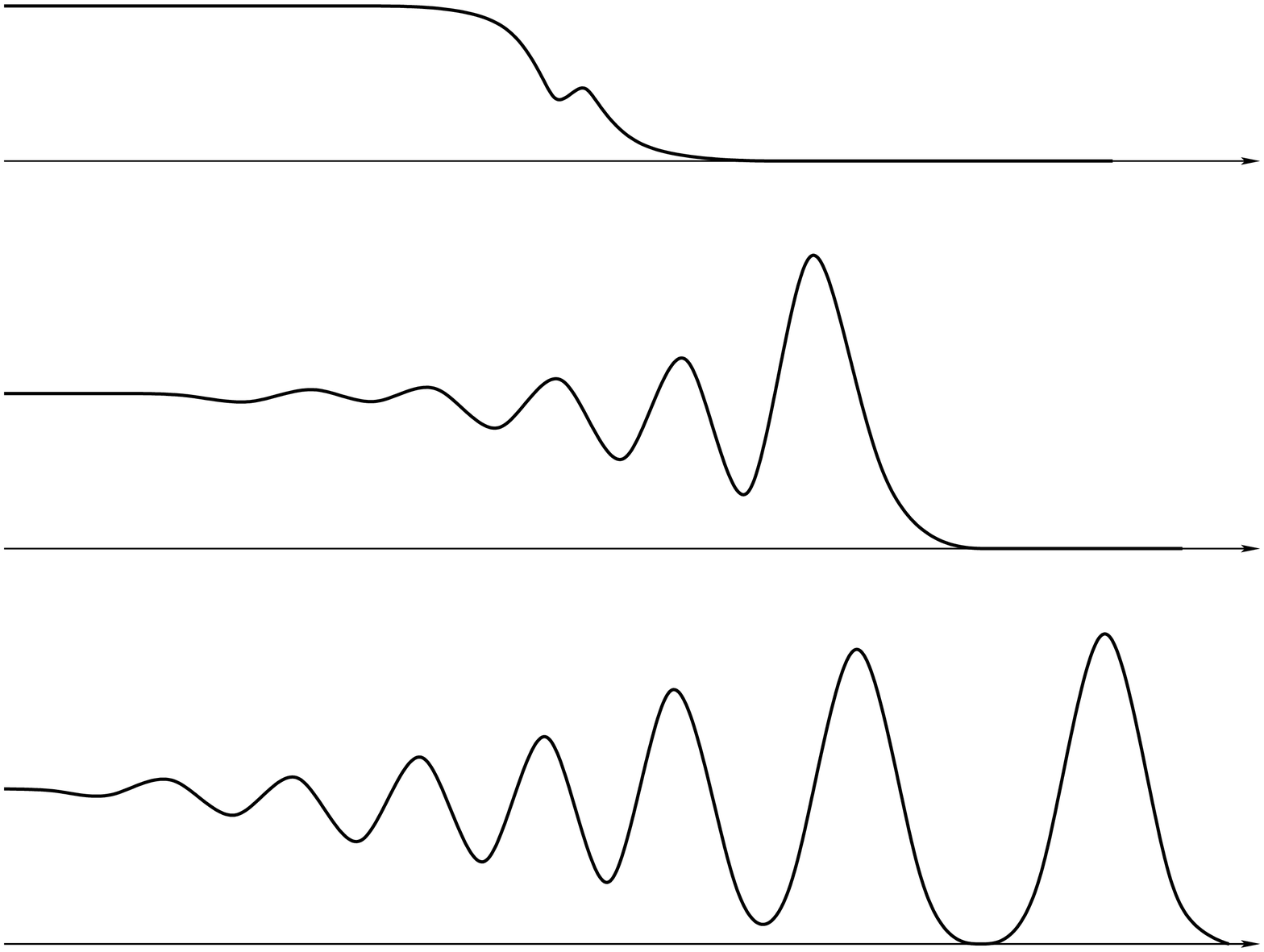}
\end{center}
\caption{The development of oscillation zone from the step-like
initial data for the case of equation (\ref{plus}).}
\label{StepOscPlus}
\end{figure}

First of all, we need to understand whether  
hydrodynamic solutions for $r_1$ and $r_2$ break down and develop 
a singularity.  For $\mu_0>0$ and the initial  density profile shown in
Fig. \ref{StepInCond}  the steepness of function
$r^{1} (X)$ should increase with time while the steepness of
function $r^{2} (X)$ should decrease with time (see Fig. \ref{r1r2Hydr}).
This follows from simple hydrodynamic analysis following
equations (\ref{r1MovSyst}) and (\ref{r2MovSyst}).
This means that the steepness of solutions $\rho (X)$, 
$k (X)$ will increase on the left-moving edge of Fig. 
\ref{TwoSteps} and decrease on the right-moving edge. (The
situation changes to the opposite for the inverse step initial
state.) So in this case no dispersive corrections are needed for
$r^{2} (X)$. Function $r^{2} (X)$ should remain smooth for all 
$T > 0$ in the hydrodynamic approximation.
On the other hand, function $r^{1} (X)$ develops a breaking point in the
hydrodynamic approximation. Thus we need to consider
equation (\ref{r1disp}) taking  into account dispersive corrections.
As we discussed before dispersive corrections should give rise 
the oscillation zone, which we expect to grow linearly with time.
The form of oscillations should be different
for  equations (\ref{minus}) and (\ref{plus}) due
to different signs of dispersion in these systems 
(see Fig. \ref{StepOscMinus} - \ref{StepOscPlus}).

To understand the oscillation zone that arises following
breaking of the hydrodynamic solution we need to analyze
dynamics of the KdV equation with step like
initial conditions. This problem was addressed  by
A.V. Gurevich and L.P. Pitaevskii (\cite{GurPit1,GurPit2}) 
using the Whitham theory of slow modulations. 
We will now summarize their key results pointing out their
implications for our system.

\begin{figure}
\begin{center}
\includegraphics[width=14.0cm,height=5cm]{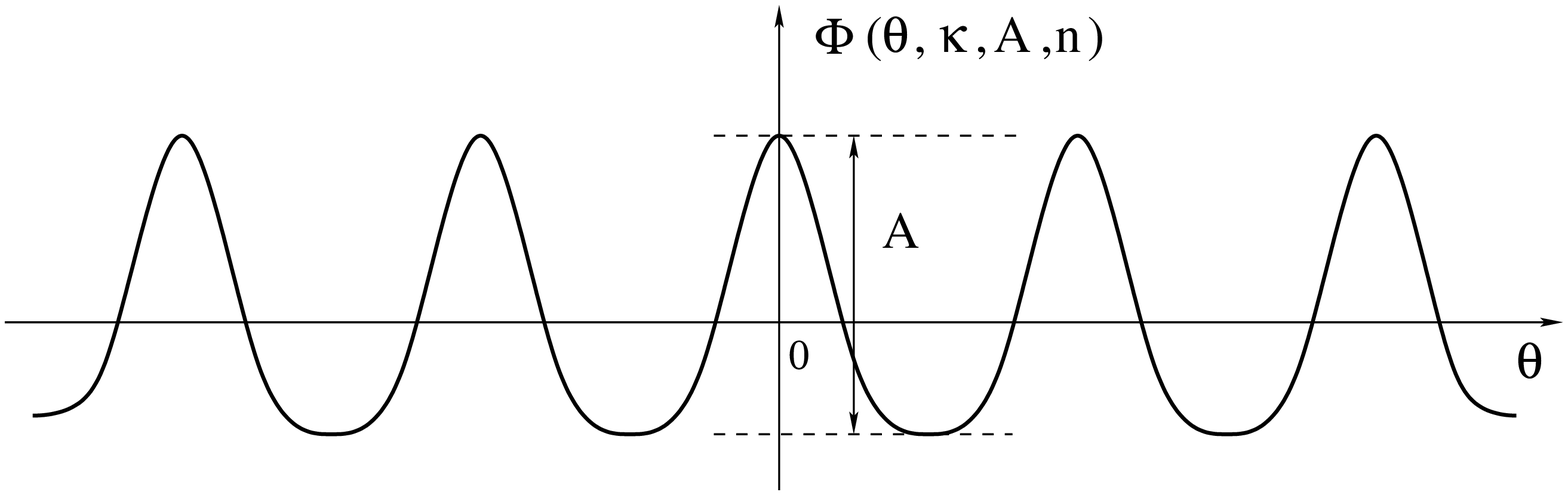}
\end{center}
\caption{The general form of the function
$\Phi (\theta, \, \kappa , A , n)$ representing the one-phase
solution of the KdV - equation.}
\label{OnePhaseSol}
\end{figure}

Gurevich and  Pitaevskii considered a general problem of slowly 
modulated one-phase solution of the KdV equation. One-phase solution
is a periodic running wave solution that provides a generalization
of the  one-soliton solutions of the KdV equation
$$U (X, T) \,\, = \,\, \Phi \left( 
\kappa X + \omega T + \theta_{0} , \, \kappa , A , n \right) $$
One-phase solution depends on three parameters $(\kappa , A , n)$. 
Functions $\Phi (\theta, \, \kappa , A , n)$ should be 
$2\pi$-periodic in $\theta$, so parameter $\kappa$
plays the role of the wave number for nonlinear running 
waves. Parameter $A$ plays the role of the amplitude of the
periodic solution, while parameter $n = <\Phi>$ is the
value of $\Phi$ averaged over one period (see Fig.
\ref{OnePhaseSol}).

One-phase
solutions of KdV can be written in the form

$$\Phi (\kappa X + \omega T, \kappa , A , n) \,\, = \,\, 
{A \over s^{2}}
\,\, {\rm dn}^{2} \left[ \left( {A \over 12 s^{2}}\right)^{1/2}
(X - V T), s \right] \, + \, \gamma $$
$$V \,\, = \,\, {A \over 3 s^{2}} (2 - s^{2}) \, + \,
\gamma $$
where $s$ is the modulus of the Jacobi elliptic function
${\rm dn} (u, s)$, $0 \leq s \leq 1$. The values 
$(\kappa, \omega, n)$ 
can be expressed in terms of the parameters $(A, s, \gamma)$ in the 
following way

$$ \kappa \, = \, {\pi \over K (s)} \left(
{A \over 12 s^{2}}\right)^{1/2} \,\,\, , \,\,\,
\omega \, = \, - \, V \, \kappa \, = \, - \,
{4 \pi \over K (s)} (2 - s^{2})
\left( {A \over 12 s^{2}}\right)^{3/2} \, - \,
{\gamma \pi \over K (s)} \left( {A \over 12 s^{2}}\right)^{1/2} $$
$$n \, = \, \gamma \, + \, {A E (s) \over s^{2} K (s)} $$   
where $K (s)$ and $E (s)$ are the elliptic integrals of the first
and the second kind respectively.

 We can also write 

$$\Phi (\theta, A, s, \gamma) \,\, = \,\,
{A \over s^{2}} \, {\rm dn}^{2} \left(
{K (s) \over \pi} \theta, s \right) \, + \, \gamma $$
as normalization of function 
$\Phi (\theta, \kappa , A , n)$.

 The one-soliton solutions of KdV can be considered as the 
limiting case of the one-phase solutions in the large-period
limit $\kappa \rightarrow 0$. Traditionally the asymptotes
$\Phi (\theta) \rightarrow 0$, $\theta \rightarrow \pm \infty$
is assumed for the soliton solutions of KdV, so the amplitude
parameter $A$ remains the one parameter of a one-soliton
solution.

In Whitham's approach parameters $(\kappa , A , n)$
become slow functions of $x$ and $t$ 

$$\kappa \, = \, \kappa (X, T) \,\,\, , \,\,\,
A \, = \, A (X, T) \,\,\, , \,\,\, n \, = \, n (X, T) $$
so that functions $\kappa (X, T)$, $A (X, T)$, $n (X, T)$
satisfy a nontrivial system of quasilinear equations in partial
derivatives (the so-called Whitham's system). Whitham's system describes
evolution of  initial parameters $\kappa (X, 0)$, $A (X, 0)$, 
$n (X, 0)$ of oscillating solutions, such that development
of oscillations can be  calculated in this case.

Gurevich and Pitaevskii showed that  
in the KdV equation with a step like initial
conditions, the small oscillation zone, that arises near 
the breaking  point of the hydrodynamic solution,
can be described by the self-similar solutions
characterized by only one variable, $l=X/T$.

 In more details, the asymptotic ($T \rightarrow \infty$) form of 
oscillations can be described by the modulated one-phase solutions of 
KdV with parameters $\kappa (X, T)$, $A (X, T)$, $n (X, T)$ of the form

$$\kappa (X, T) \, = \, \kappa (X/T) \,\,\, , \,\,\,
A (X, T) \, = \, A (X/T) \,\,\, , \,\,\,
n (X, T) \, = \, n (X/T) $$

 The oscillation zone is located in the interval

$$l_{-} \,\, < \,\, X / T \,\, < \,\, l_{+} $$
in this asymptotic regime.

 According to \cite{GurPit1} - \cite{GurPit2} the amplitude
of oscillations $A (X, T)$ becomes zero at the "trailing edge"
of the oscillation zone (the right edge in Fig. \ref{StepOscMinus}
and the left edge in Fig. \ref{StepOscPlus}). The wave number
of nonlinear oscillations $\kappa (X, T)$ becomes zero at the
"leading edge" of the oscillation zone (the left edge in Fig.
\ref{StepOscMinus} and the right edge in Fig. \ref{StepOscMinus}).

 We can see that the "trailing edge" of oscillation zone can 
be considered as a source of oscillations with small amplitude,
which develop into solitons in the limit $T \rightarrow \infty$.
The "leading edge" of the oscillation zone can be considered as
a source of free solitons since we have 
$\kappa \rightarrow 0$ on this edge and the distance between 
solitons tends to infinity for $T \rightarrow \infty$.

We point out  that it is also possible to analyze the problem
above  in terms of the "pure" soliton
picture (\cite{LaxLev,LaxLevVen}). Approach used in
(\cite{LaxLev,LaxLevVen}) is also a classical part
of the soliton theory.

General problem of the decay of different initial configurations 
in the theory of  small-dispersion KdV-equation represents a big 
branch of the soliton theory. While we do not 
discuss other problems here, we expect that many of the known 
mathematical results will be relevant for different experiments
with ultracold atoms. We also point out that  our  methodology for 
identifying the character of solitons 
(particle- or hole-like) was based on considering
the function $U (X, T)$, which describes
Riemann invariants  $r^{1} (X, T)$ or $r^{2} (X, T)$. 
It is more natural to classify solitons based on the density. 
Relations between $r^{\{1,2\}} (X, T)$ and the more physical 
variables of the density, $\rho (X, T)$, and the phase gradient, 
$k (X, T)$, are given in equation (\ref{rhokExp}).
We find that the density always follows the behavior of 
$r^{\{1,2\}} (X, T)$. Hence our classification of the hole-type 
and the particle-type solitons in terms of the density coincides 
with that given in terms of the function $U (X, T)$.

Before concluding this section we would like to point
out that whether step-like conditions shown in
Fig. \ref{StepInCond} should be considered as a source 
of hole-like or particle-like solitons in the solutions
$\rho (X, T)$, $k (X, T)$ depends on the relation between
parameters $(J_{\perp}, J_{z}, \mu_{0})$. In general, we expect 
that larger values of $J_{z}$ and $\mu_{0}$ suppress
the appearance of hole-type solitons and favor 
solitons of the particle type. On the opposite side, smaller
values of $J_{z}$ and $\mu_{0}$ allow solitons of the
hole type and suppress solitons of
particle type.

\section{Two-dimensional effects.} 
\label{2DSection}

In this section we discuss the role of transverse directions.
We consider a question of whether one dimensional profiles, that we 
discussed so far, are stable against "weak" modulation in the transverse 
direction.
 
 For a $D$-dimensional lattice
we need to change the long wavelength Lagrangian density
(\ref{ThetaSigmaLagr}) to a more general expression

$${\cal L} \,\, = \,\, {1 \over 2} \, \sigma_{T} \, \cos \theta
\, - \, 2 \, J_{\perp} \, \sin^{2} \theta \,
\sum_{i=1}^{D} \cos \sigma_{X^{i}}
\, - \, 2 \, J_{z} \, D \, \cos^{2} \theta \, - $$
$$- \, h^{2} \, J_{\perp} \, \sin \theta \, \sum_{i=1}^{D}
\left( \sin \theta \right)_{X^{i}X^{i}}
\, \cos \sigma_{X^{i}} \, - \, h^{2} \, J_{z} \, \cos \theta \,
\sum_{i=1}^{D} \left( \cos \theta \right)_{X^{i}X^{i}} \, - $$
$$- \, h^{2} \, J_{\perp} \, \sin^{2} \theta \, \sum_{i=1}^{D}
\left( {1 \over 6} \, \sigma_{X^{i}X^{i}X^{i}} \, \sin \sigma_{X^{i}}
\, + \, {1 \over 4} \, \sigma_{X^{i}X^{i}}^{2} \, \cos \sigma_{X^{i}}
\right) \,\, + \,\, {\cal O} (h^{4}) $$
or
 
$${\cal L} \,\, = \,\, {1 \over 2} \, \sigma_{T} \, \mu \, - \,
2 \, J_{\perp} \, (1 - \mu^{2}) \, \sum_{i=1}^{D}
\cos \sigma_{X^{i}} \, - \, 2 \, J_{z} \, D \, \mu^{2} \, + $$
$$+ \, h^{2} \, J_{\perp} \,
{\mu^{2} \over 1 - \mu^{2}} \,
\sum_{i=1}^{D} \mu_{X^{i}}^{2} \cos \sigma_{X^{i}} \, - \, 
h^{2} \, J_{z} \, \mu \, \sum_{i=1}^{D} \mu_{X^{i}X^{i}} \, + $$
$$+ \, h^{2} \, J_{\perp} \, (1 - \mu^{2}) \, \sum_{i=1}^{D}
\left( {1 \over 3} \, \sigma_{X^{i}X^{i}X^{i}} \,
\sin \sigma_{X^{i}} \, + \, {1 \over 4} \, \sigma_{X^{i}X^{i}}^{2}
\, \cos \sigma_{X^{i}} \right) \,\, + \,\, {\cal O} (h^{4}) $$
in the coordinates $(\mu, \sigma)$.

\begin{figure}   
\begin{center}
\includegraphics[width=14.0cm,height=8cm]{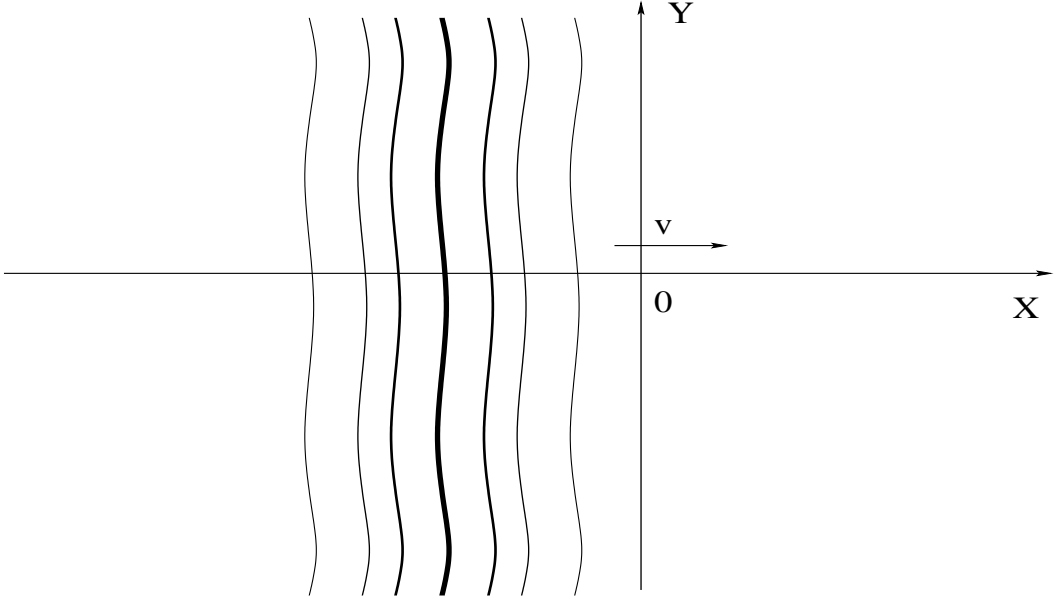}
\end{center}
\caption{Schematic sketch of a one-dimensional soliton string
modulated in the $Y$-direction.}
\label{SolTwoDimMod}
\end{figure}

We separate  the hydrodynamic and  dispersive parts
of the Lagrangian and repeat considerations used in the
previous sections. Analysis of the  dynamical
system is more complicated for $D > 1$ and we will not
explore all of its richness. We only address a question whether
one dimensional solitons that we discussed so far are stable
with respect to formation of a two dimensional pattern  
(see Fig. \ref{SolTwoDimMod}).

 We start with a generic situation corresponding to equation
(\ref{minus}) or (\ref{plus}). Since we are going to consider only
small modulations of the soliton strings, we can follow the
procedure suggested in \cite{KadPet} to get the 
Kadomtsev - Petviashvili equation for two-dimensional systems.

 Firstly we  recall  that equations (\ref{r1disp})-(\ref{r2disp})
are written in the moving coordinate systems. For
the left-moving part of the solution in
the laboratory frame of reference we have

$$r^{1}_{T} \,\, = \,\, \sqrt{J_{\perp} (J_{\perp} - J_{z})} \,
\sqrt{32 ( 1 - \mu_{0}^{2} )} \,\, r^{1}_{X}
\, - \, 6 \, \mu_{0} \,
\sqrt{J_{\perp} (J_{\perp} - J_{z})} \,\, r^{1} \,\, r^{1}_{X} \, + $$
\begin{equation}
\label{r1LabSyst}
+ \, \sqrt{2} \, h^{2} \, \sqrt{1 - \mu_{0}^{2}} \,
\sqrt{{J_{\perp} \over J_{\perp} - J_{z}}} \,
\left( J_{\perp} \, \left( {1 \over 6} \, - \,
{\mu_{0}^{2} \over 1 - \mu_{0}^{2}} \right) \, - \,
{7 \over 6} \, J_{z} \right) \, r^{1}_{XXX} 
\end{equation}

In writing the last equation  we preserved the restriction
$r^{2} = const$

 The first term in the right-hand part plays  the main role in
the evolution of $r^{1}(X,T)$ and  other terms represent 
small corrections with respect to the main contribution. According
to \cite{KadPet} we only need to calculate corrections to the main
term coming from the slow modulation of the solution in the $Y$-direction.
This procedure gives us stable or unstable variants of the Kadomtsev - 
Petviashvili equation. The main term in the right-hand part of
(\ref{r1LabSyst}) originates from the linear system 
(\ref{LinkmuSyst}) of (\ref{LinrhoSyst}) which can be easily
written in the two-dimensional form by adding additional
derivatives in the $Y$-direction. 
What we need here is
correction to the dispersion law $\omega^{2} \sim k_{X}^{2}$
which can be written as

$$\omega \,\, = \,\, v_{0} \, \sqrt{k_{X}^{2} + k_{Y}^{2}}
\,\, \sim \,\, \sqrt{J_{\perp} (J_{\perp} - J_{z})} \,
\sqrt{32 ( 1 - \mu_{0}^{2} )} \,\, \left( k_{X} \, + \,
{1 \over 2} \, {k_{Y}^{2} \over k_{X}} \right) $$
for the left-moving part in our situation\footnote{We use
the expansion of the solutions of linear system in the form
$f (X,T) = \int f({\bf k}) e^{i \omega ({\bf k}) T + i {\bf k} {\bf R}} 
d {\bf k}$.}. As a result, the small modulations in the 
$Y$-direction of solutions of (\ref{r1LabSyst})
can be described by the equation

$$r^{1}_{TX} \,\, = \,\, \sqrt{J_{\perp} (J_{\perp} - J_{z})} \,
\sqrt{32 ( 1 - \mu_{0}^{2} )} \,\, r^{1}_{XX}
\, - \, 6 \, \mu_{0} \,
\sqrt{J_{\perp} (J_{\perp} - J_{z})} \,\, 
\left( r^{1} \, r^{1}_{X} \right)_{X} \, + $$
$$ + \, \sqrt{2} \, h^{2} \, \sqrt{1 - \mu_{0}^{2}} \,
\sqrt{{J_{\perp} \over J_{\perp} - J_{z}}} \,
\left( J_{\perp} \, \left( {1 \over 6} \, - \,
{\mu_{0}^{2} \over 1 - \mu_{0}^{2}} \right) \, - \,
{7 \over 6} \, J_{z} \right) \, r^{1}_{XXXX} \, + $$
$$+ \, {1 \over 2} \, \sqrt{J_{\perp} (J_{\perp} - J_{z})} \,
\sqrt{32 ( 1 - \mu_{0}^{2} )} \,\, r^{1}_{YY} $$
or

$$r^{1}_{TX} \,\, = \,\,  {1 \over 2} \, 
\sqrt{J_{\perp} (J_{\perp} - J_{z})} \,
\sqrt{32 ( 1 - \mu_{0}^{2} )} \,\, r^{1}_{YY}
\, - \, 6 \, \mu_{0} \,
\sqrt{J_{\perp} (J_{\perp} - J_{z})} \,\,
\left( r^{1} \, r^{1}_{X} \right)_{X} \, + $$
\begin{equation}
\label{KadomPetv}
 + \, \sqrt{2} \, h^{2} \, \sqrt{1 - \mu_{0}^{2}} \,
\sqrt{{J_{\perp} \over J_{\perp} - J_{z}}} \,
\left( J_{\perp} \, \left( {1 \over 6} \, - \,
{\mu_{0}^{2} \over 1 - \mu_{0}^{2}} \right) \, - \,
{7 \over 6} \, J_{z} \right) \, r^{1}_{XXXX} 
\end{equation}
in the moving coordinate system.

 Equation (\ref{KadomPetv}) is the Kadomtsev - Petviashvili
(KP) equation which describes the small transverse modulations 
of solutions of the KdV equation considered in the 
two-dimensional case. The stable Kadomtsev - Petviashvili
equation corresponds to the same signs of the coefficients
for $r^{1}_{XXXX}$ and $r^{1}_{YY}$. In this case the small
modulation of a soliton string causes just the weak
oscillations along the string and does not produce any
instability. The opposite situation with different signs of
the coefficients before $r^{1}_{XXXX}$ and $r^{1}_{YY}$
corresponds to the unstable situation where the soliton
strings are unstable with respect to modulation along the
$Y$-axis.

 We can see then that the stable soliton string in two
dimensions arises for the situation of equation (\ref{minus}),
i.e.

$$J_{z} \, < \, J_{\perp} / 7 \,\,\,\,\,\,\,\, , \,\,\,\,\,\,\,\,
- \sqrt{{J_{\perp} - 7 J_{z} \over 7 (J_{\perp} - J_{z})}}
\,\, < \,\, \mu_{0} \,\, < \,\,
\sqrt{{J_{\perp} - 7 J_{z} \over 7 (J_{\perp} - J_{z})}} $$
which corresponds to the small values of $J_{z}$ and the
density $n \sim 1/2$ in the pattern.

 The solutions we considered in the opposite situation

$$J_{z} \, > \, J_{\perp} / 7 \,\,\,\,\,\,\,\, {\rm or}
\,\,\,\,\,\,\,\, |\mu_{0}| \,\, > \,\, 
\sqrt{{J_{\perp} - 7 J_{z} \over 7 (J_{\perp} - J_{z})}} $$
are unstable from the point of view of the two-dimensional
modulations.\footnote{Let us note here that these conclusions
do not require in fact the square two-dimensional lattice
and are applicable for any dispersion law
$\omega^{2} = \alpha k_{X}^{2} + \beta k_{Y}^{2}$,
$\alpha, \beta > 0$ in the main linear approximation.}

 Let us say now that the analogous considerations can be performed
also in the case of equation (\ref{mKdV}) so the results formulated
above can be used also in the limit $\mu_{0} \rightarrow 0$.

 We must certainly say that the Kadomtsev - Petviashvili
equation is an integrable system from the point of view of
the inverse scattering methods (\cite{Druma,ZakhShab2}).
The theory of equation (\ref{KadomPetv}) is very deep and brought
many beautiful ideas in the theory of solitons. Let us just mention
here two nice classes of solutions of (\ref{KadomPetv}) in
the stable and the unstable situation.

\vspace{0.5cm}

\noindent
1) The most interesting solutions of the Kadomtsev - Petviashvili
equation in the stable situation are the two-dimensional
$N$-soliton solutions which are described in general by the
formula

$$U (X, Y, T) \,\, = \,\, \Phi \left( \omega^{1} T +
k^{1}_{X} X + k^{1}_{Y} Y + c^{1}, \dots, \omega^{N} T +
k^{N}_{X} X + k^{N}_{Y} Y + c^{N} \right) $$
with some special functions 
$\Phi (\theta^{1}, \dots, \theta^{N})$ (\cite{Satsuma}).

 The $N$-solution solutions of the KP equation represent $N$
plane interacting waves propagating at some angles with respect
to each other. The interaction of the waves results in the phase
shifts which can be rather big in the resonant case (\cite{Miles}).

\vspace{0.5cm}

\noindent
2) For the unstable variant of the KP equation very interesting
rational localized solutions ("lumps") can arise. The "lumps"
represent localized both in $X-$ and $Y-$direction solitons with 
rational dependence of coordinates. The interaction of solitons
does not produce any phase shifts in this situation, so the
solitons completely "forget" about each other after the interaction
(\cite{BIMMZ}).

\vspace{0.5cm}

 Let us emphasize here that the relation $J_{\perp} > J_{z}$
was assumed everywhere in our considerations above and the
properties we consider will be completely changed for the
opposite situation $J_{\perp} < J_{z}$. Thus, as we pointed
out already, the hydrodynamic approximation (\ref{kmuHydrSyst})
reveals an elliptic instability for the small values of $k$
($k < \pi/2$) in this situation which corresponds to a modulation
instability of long-wave solutions of (\ref{DiscreteSystem})
in this case. In the same way, equation (\ref{NonLinShr})
becomes the focusing nonlinear Shr\"odinger equation in this
situation which corresponds to the unstable behavior of the
long-wave solutions of (\ref{DiscreteSystem}) either.
However, the integrable nature of the focusing nonlinear
Shr\"odinger equation leads to very interesting behavior of
solutions also in this case. The most interesting part is the
presence of the $N$-soliton solutions for the focusing NLS
equation which should be observed for $J_{z} > J_{\perp}$.
The corresponding two-dimensional equation for (\ref{NonLinShr}) 
can be written in the form

\begin{equation}
\label{NLSTwoDim}
i h \, \psi_{T} \,\, = \,\, 8 \, \left( J_{\perp} - J_{z} \right)
\, \psi \, - \, 4 \, \left( J_{\perp} - J_{z} \right) \,    
|\psi|^{2} \psi \, + \, 4 \, h^{2} \, J_{\perp} \, \psi_{XX}
\, + \, 4 \, h^{2} \, J_{\perp} \, \psi_{YY}
\end{equation}

The one-dimensional solutions of (\ref{NLSTwoDim}), however,
are unstable with respect to the weak transverse modulations
(\cite{BespTal}) for $J_{z} > J_{\perp}$.

\section{Concluding remarks}

 Soliton solutions in quantum systems
is  a subject of considerable theoretical interest. However, most of the 
earlier work  focused on one dimensional systems, where special analytical 
tools, such as the Bethe ansatz solution, are available. For example, 
exact solitonic solutions were considered recently in a different quantum 
system in a series of papers \cite{BetAbWieg1,BetAbWieg2,BetAbWieg3}. 
Their analysis relied on the quantum inverse scattering methods, which 
are special to 1d integrable systems. Our analysis in this paper is on 
constructing semiclassical solitons in two and three dimensional systems. 

States described by the wavefunction (\ref{WaveFunction}) correspond
to collective excitations in the superfluid state. In the superfluid state
the U(1) symmetry is spontaneously broken, so the number
of particles is not a good quantum number. Solitons
which we discuss in this papers are semiclassical collective excitations. 
They can be thought
 of as spatially inhomogeneous coherent states representing
 non-linear excitations of the Hamiltonian.
 These solitons do not have a well defined number of particles.  
Within our approximations solitons have infinite lifetime.
We expect that including coupling to other excitations may give rise
to small but finite decay rate for the solitons, which may lead to 
dissipative terms in the semiclassical dynamics. We expect  that this
should not change our conclusions qualitatively, since solitons
should be robust against small dissipation \cite{Naumkin1991}.

\section{Acknowledgments}

\vspace{0.5cm}

We thank B. Altshuler, 
I. Bloch, M. Greiner, B. Halperin,
M. Lewenstein,   D. Pekker, and G. Refael for insightful discussions.
This work was partially supported (E.D.) by the NSF Grant No. 
DMR-07-05472, DARPA OLE program, CUA, AFOSR Quantum Simulation MURI, 
AFOSR MURI on Ultracold Molecules, the ARO-MURI on Atomtronics.
We also acknowledge support from the Harvard ITAMP.

\section{Appendices}

\subsection{General approach for analyzing  solitonic solutions in 
KdV-type equations}
\label{appendix_kdv}

The famous procedure of integration of the KdV equation
(\cite{GGKM}) is based on the connection of the KdV with the linear
Shr\"odinger operator passing through the iso-spectral deformations
according to the KdV evolution. The corresponding linear problems
have the form

\begin{equation}
\label{MinusLinProb}
- \, \psi_{XX} \, + \, U \, \psi \,\, = \,\, E \, \psi
\end{equation}
for equation (\ref{minus}), and

\begin{equation}
\label{PlusLinProb}
- \, \psi_{XX} \, - \, U \, \psi \,\, = \,\, E \, \psi
\end{equation}
for equation (\ref{plus}). The connection of the KdV equations
with the linear problems (\ref{MinusLinProb}) - (\ref{PlusLinProb})
gives a possibility to represent also equations 
(\ref{minus}) - (\ref{plus}) in the equivalent form (\cite{Lax}):

\begin{equation}
\label{LaxRepr}
{\partial \over \partial T} \, {\hat L} \,\, = \,\,
{\hat L} \, {\hat A} \,\, - \,\, {\hat A} \, {\hat L}
\end{equation}
where the operators ${\hat L}$, ${\hat A}$ have the form

$${\hat L} \,\, = \,\, - \, {d^{2} \over d X^{2}} \,\, + \,\, U
\,\,\,\,\,\,\,\, , \,\,\,\,\,\,\,\,
{\hat A} \,\, = \,\, - \, 4 \, {d^{3} \over d X^{3}} \, + \,
6 \, U \, {d \over d X} \, + \, 3 \, U_{X} $$
for equation (\ref{minus}) and

$${\hat L} \,\, = \,\, - \, {d^{2} \over d X^{2}} \,\, - \,\, U
\,\,\,\,\,\,\,\, , \,\,\,\,\,\,\,\,
{\hat A} \,\, = \,\,  4 \, {d^{3} \over d X^{3}} \, + \,
6 \, U \, {d \over d X} \, + \, 3 \, U_{X} $$
for equation (\ref{plus}). Representation (\ref{LaxRepr}) of the
KdV equation permits to consider the KdV evolution as the 
isospectral deformation of the operator ${\hat L}$ using the
exponent of the operator ${\hat A}$ as the corresponding basis
transformation.

 According to the procedure represented in \cite{GGKM} the
scattering problem for the linear equations (\ref{MinusLinProb})
and (\ref{PlusLinProb}) plays the basic role in solving equations
(\ref{minus}) and (\ref{plus}) in the rapidly decreasing case
$|U (X)| \rightarrow 0$, $X \rightarrow \pm \infty$. Thus, if we
consider the eigen-functions of (\ref{MinusLinProb}) or
(\ref{PlusLinProb}) having the asymptotic form

$$\psi (X) \,\, \simeq \,\, e^{ikX} \,\, + \,\, b (k) \,
e^{-ikX} \,\,\,\,\, , \,\,\,\,\,
X \rightarrow - \infty
\,\,\,\,\,\,\,\,\,\, , \,\,\,\,\,\,\,\,\,\,
\psi (X) \,\, \simeq \,\, a (k) \, e^{ikX}
\,\,\,\,\, , \,\,\,\,\,
X \rightarrow \infty $$
($k^{2} = E$) and introduce the reflection and transition
coefficients $r (k)$, $t (k)$ in the standard way we will have
very simple evolution of the functions $r (k, T)$, $t (k, T)$:

$$t (k, T) \,\, = \,\, t (k, 0)
\,\,\,\,\,\,\,\, , \,\,\,\,\,\,\,\,
r (k, T) \,\, = \,\, e^{\pm 8ik^{3}T} \,\, r (k, 0) $$
according to the KdV evolution of $U (X, T)$.\footnote{We have
different signs in the evolution of $r$ for equations
(\ref{minus}) and (\ref{plus}).}

 In the same way, if the potential $U (X)$ has bounded states
$\psi_{n} (X)$ with the energies $E_{n}$ we will have
$E_{n} = const$ during all the KdV evolution. From the other
hand, provided that the functions $\psi_{n} (X)$ are 
normalized in the following way

$$\psi_{n} (X) \,\, \simeq \,\, e^{k_{n} X}
\,\,\,\,\, , \,\,\,\,\, X \rightarrow - \infty
\,\,\,\,\,\,\,\,\,\, , \,\,\,\,\,\,\,\,\,\,
\psi_{n} (X) \,\, \simeq \,\, C_{n} \, e^{- k_{n} X}
\,\,\,\,\, , \,\,\,\,\,
X \rightarrow \infty $$
($- k_{n}^{2} = E_{n}$) the evolution of the values $C_{n}(T)$ 
is given by $C_{n}(T) = e^{\pm 8 k_{n}^{3} T} \, C_{n}(0)$.

 The full set of the scattering data

$$\{ r (k) , E_{n} , C_{n} \} $$
gives the full information about the potential $U (X)$
(\cite{GelLev,Mar,KayMos}) such that the solution $U (X, T)$
can be reconstructed at every time $T$ using the values of
$r (k, T)$, $E_{n}$, $C_{n}(T)$.

 The potentials $U (X)$ having zero reflection coefficient
$r (k) \equiv 0$ are called the reflectionless potentials
and correspond to the exact $N$-soliton solutions of the
KdV-equation. The number of the bounded states 
($n = 1, \dots, N$) is equal to the number of solitons in the
$N$-soliton solution, so we can say that every bounded state
in potential $U (X)$ corresponds to a soliton in the solution
$U (X, T)$. The one-soliton solutions of the KdV-equation
have the form

\begin{equation}
\label{onesolminus}
U (X, T) \,\, = \,\, - \, 
{2 a^{2} \over {\rm ch}^{2} (a X + 4 a^{3} T + c_{0})}
\end{equation}
for equation (\ref{minus}) and

\begin{equation}
\label{onesolplus}
U (X, T) \,\, = \,\, 
{2 a^{2} \over {\rm ch}^{2} (a X - 4 a^{3} T + c_{0})}
\end{equation}
for equation (\ref{plus}). Potentials (\ref{onesolminus}) and
(\ref{onesolplus}) have exactly one bounded state according
to linear problems (\ref{MinusLinProb}) and (\ref{PlusLinProb})
respectively with energy $E_{1} = E_{1} (a)$ depending on the
amplitude of a soliton.

\subsection{Analysis of solitons close to half-filling. Modified KdV 
equation}

In this section we discuss  soliton solutions of two types
of the mKdV equations :

\begin{equation}
\label{MinusmKdV}
U_{T} \, + \, 6 \, U^{2}\, U_{X} \, - \, U_{XXX} \,\, = \,\, 0
\end{equation}
\begin{equation}
\label{PlusmKdV}
U_{T} \, + \, 6 \, U^{2}\, U_{X} \, + \, U_{XXX} \,\, = \,\, 0
\end{equation}
(we put $\alpha = 0$ here).

\begin{figure}
\begin{center}
\includegraphics[width=14.0cm,height=5cm]{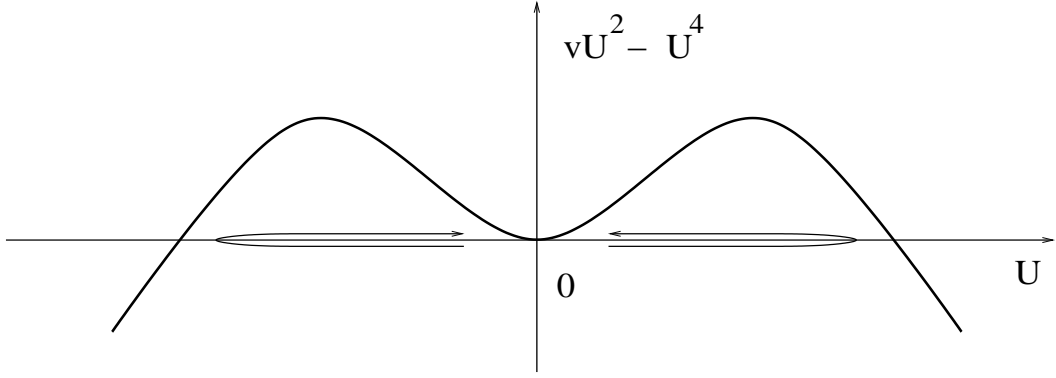}
\end{center}
\caption{The left and right paths of integration w.r.t. $U$
corresponding to hole- and particle-type one-soliton
solutions of (\ref{PlusmKdV}).}
\label{mKdVPlus}
\end{figure}

Equation (\ref{PlusmKdV})
has two varieties of one-soliton solutions of arbitrary amplitude
defined by the analytic formula

$$\pm \,\, \int {d U \over \sqrt{v U^{2} - U^{4}}} \,\, = \,\,
X \, + \, C $$
Here  $U$ should be taken from one of the regions in the 
$U$-space where the value of expression $v U^{2} - U^{4}$
is positive (see Fig. \ref{mKdVPlus}).

\begin{figure}
\begin{center}
\includegraphics[width=14.0cm,height=6cm]{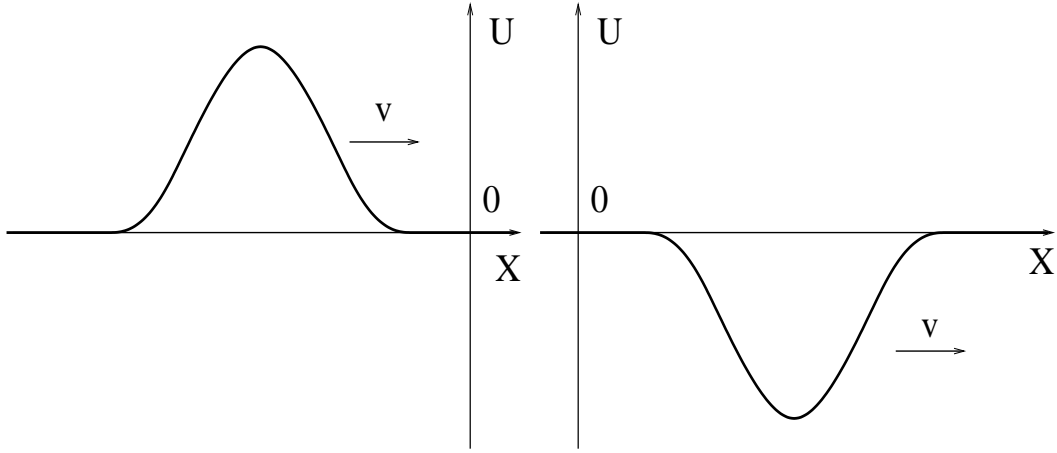}
\end{center}
\caption{The particle- and the hole-type solitons for
equation (\ref{PlusmKdV}).}
\label{PlusmKdVSol}
\end{figure}

 It is then easy to see that we can have either the particle-type 
or hole-type solitons, both moving to the right ($v > 0$)
\footnote{We remind the readers that this analysis is done in the 
left-moving coordinate system. Velocity of solitons with respect 
to the moving frame should be much smaller than the velocity of the 
reference frame moving.} and connected by the transformation 
$U \rightarrow - U$ (Fig. \ref{PlusmKdVSol}).

 Soliton velocity is proportional to the square of the 
amplitude $v \sim A^{2}$ and we can have arbitrary positive value 
of $A$. Explicit formula for the one-soliton solutions of
(\ref{PlusmKdV}) can be written in the form

$$U \,\, = \,\, \pm \,\, 
{a \over {\rm ch} \, (a X - a^{3} T + c_{0})} $$

\begin{figure}
\begin{center}
\includegraphics[width=14.0cm,height=6cm]{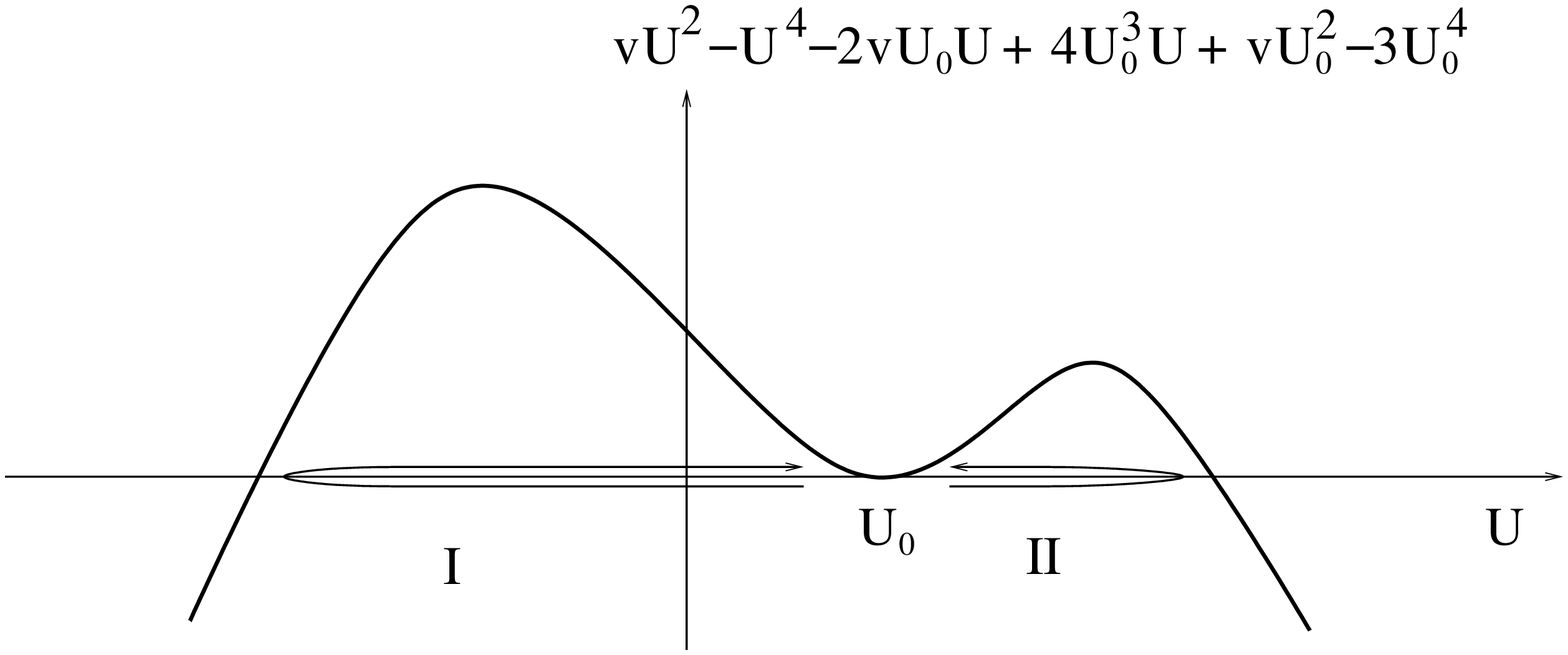}
\end{center}
\caption{The left and right paths of integration w.r.t. $U$
corresponding to the hole- and t particle-type soliton
solutions on a "pedestal" $U_{0}$ for (\ref{PlusmKdV}).}
\label{PlusSolPed}
\end{figure}

Equation (\ref{PlusmKdV}) also admits
more general soliton solutions. One can construct
soliton solutions on a "pedestal" . These solutions are defined
by a more general analytic formula

$$\pm \,\, \int 
{d U \over \sqrt{v U^{2} - U^{4} - 2 v U_{0} U + 4 U_{0}^{3} U
+ v U_{0}^{2} - 3 U_{0}^{4}}} \,\, = \,\, X \, + \, C $$
where two different paths of integration w.r.t. $U$ are shown
at Fig. \ref{PlusSolPed}.

 Again we can have solitons of the particle and hole type, both 
on a "pedestal" $U = U_{0}$
moving with the speed $v$ (our discussion is done in the moving frame)
which can be represented by the
following explicit formulas

$$U \,\, = \,\, U_{0} \,\, + \,\,
{a^{2} \over \sqrt{4 U_{0}^{2} + a^{2}} \,\,
{\rm ch} \, (a X - (6 U_{0}^{2} a + a^{3}) T + c_{0})
\, + \, 2 U_{0}} $$

$$U \,\, = \,\, U_{0} \,\, - \,\,
{a^{2} \over \sqrt{4 U_{0}^{2} + a^{2}} \,\,
{\rm ch} \, (a X - (6 U_{0}^{2} a + a^{3}) T + c_{0})
\, - \, 2 U_{0}} $$

\begin{figure}
\begin{center}
\includegraphics[width=14.0cm,height=6cm]{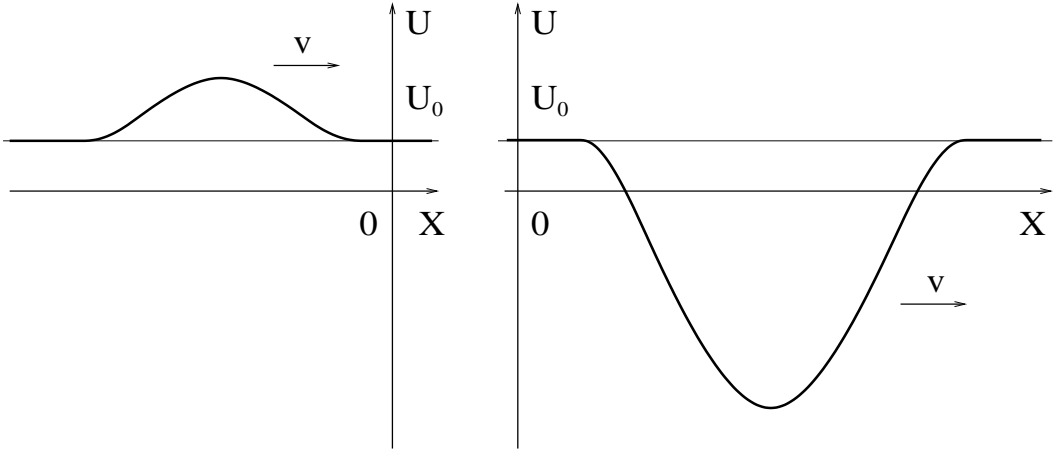}
\end{center}
\caption{Solitons of the particle- and hole- type on
the "pedestal" $U = U_{0} > 0$ moving with the same velocity $v$
for equation (\ref{PlusmKdV}).}
\label{PlusParHolPed}
\end{figure}

 We have here $v = 6 U_{0}^{2} + a^{2}$ while the amplitudes of
the particle-type and the hole-type solitons are given by the
formulas

$$A_{p.t.} \,\, = \,\, 
{a^{2} \over \sqrt{4 U_{0}^{2} + a^{2}} + 2 U_{0}} 
\,\,\,\,\,\,\,\,\, , \,\,\,\,\,\,\,\,
A_{h.t.} \,\, = \,\,
{a^{2} \over \sqrt{4 U_{0}^{2} + a^{2}} - 2 U_{0}} $$
(see Fig. \ref{PlusParHolPed}).

We can now see the difference in the particle- and hole-type
solitons in this new situation. For $U_{0} > 0$ the
amplitude of a particle type soliton can be arbitrarily small
for $a \rightarrow 0$, while the amplitude of the hole-type
soliton is bounded from below by the value $4 U_{0}$ (the
situation is opposite for $U_{0} < 0$). We  also see that
solutions, which we consider, can be described as ordinary
solitons of equation

$$U_{T} \, + \, \left( 6 \, U_{0}^{2} \, + \,
12 \, U_{0} \, U \, + \, 6 \, U^{2} \right) U_{X}
\, + \, U_{XXX} \,\, = \,\, 0 $$
after the shift $U \rightarrow U - U_{0}$. This coincides with
the general mKdV equation (\ref{mKdV}) after a Galilean
transformation.

 We can claim then that regimes described by equation
(\ref{plus}) (i.e. $J_{\perp} < 7 J_{z}$, or
$\mu_{0}^{2} > (J_{\perp} - 7 J_{z}) / 7 (J_{\perp} - J_{z})$
if $J_{\perp} > 7 J_{z}$) admit hole-type solitons after
including the next nonlinear corrections. However, the small
amplitude limit $A \rightarrow 0$ is possible only for
$\mu_{0} \rightarrow 0$ for the hole-type solutions. As a result, 
we expect that new solutions, which we discussed above, can only be 
observed  when

$$J_{\perp} \, < \, 7 J_{z} \,\,\,\,\, , \,\,\,\,\,
\mu_{0} \rightarrow 0 $$
and where changing from (\ref{plus}) to
(\ref{PlusmKdV}) is quite natural. In the regime

$$\mu_{0}^{2} \, > \, 
(J_{\perp} - 7 J_{z}) / 7 (J_{\perp} - J_{z}) $$
it is easy to see that the limit 
$\mu_{0} \rightarrow 0$ is possible  only for 
$J_{\perp} \sim 7 J_{z}$. However, as we pointed out already,
this situation is more complicated and should not be considered
from the point of view of equations (\ref{plus}) or
(\ref{PlusmKdV}). Thus, we can see that 
hole-type solitons can arise in the regimes corresponding
to equation (\ref{plus}) for the situation $J_{\perp} < 7 J_{z}$
in the limit $\mu_{0} \rightarrow + 0$
as a "reminiscent" of the region $\mu_{0} < 0$ as follows
from the higher corrections to (\ref{plus}).

 The $N$-soliton solutions of equation (\ref{PlusmKdV}) as well
as the solution of the initial value problem can be constructed
in the form analogous to the case of KdV 
(see \cite{MaxRed,Wadati}).

 We can see then that equation (\ref{PlusmKdV}) gives a good limiting 
case of equation (\ref{plus}) for $\mu_{0} \rightarrow 0$ in the 
situation $J_{\perp} < 7 J_{z}$. Moreover, equation (\ref{PlusmKdV})
provides a good limit for both cases $\mu_{0} > 0$ and
$\mu_{0} < 0$. The most remarkable feature
of this regime is that both particle- and hole-type solitons
with small amplitudes can coexist.
The cubic nonlinear correction preserves the  property of integrability
of the corresponding evolution. Hence we expect that
our analysis is applicable in the vicinity of the point
$\mu_{0} = 0$.

\vspace{0.5cm}

 Let us turn now to the regimes described by equation (\ref{minus})
(i.e. $\mu_{0}^{2} < (J_{\perp} - 7 J_{z}) / 7 (J_{\perp} - J_{z})$,
$J_{\perp} > 7 J_{z}$) which correspond to equation (\ref{MinusmKdV})
for $\mu_{0} = 0$.

\begin{figure}
\begin{center}
\includegraphics[width=14.0cm,height=6cm]{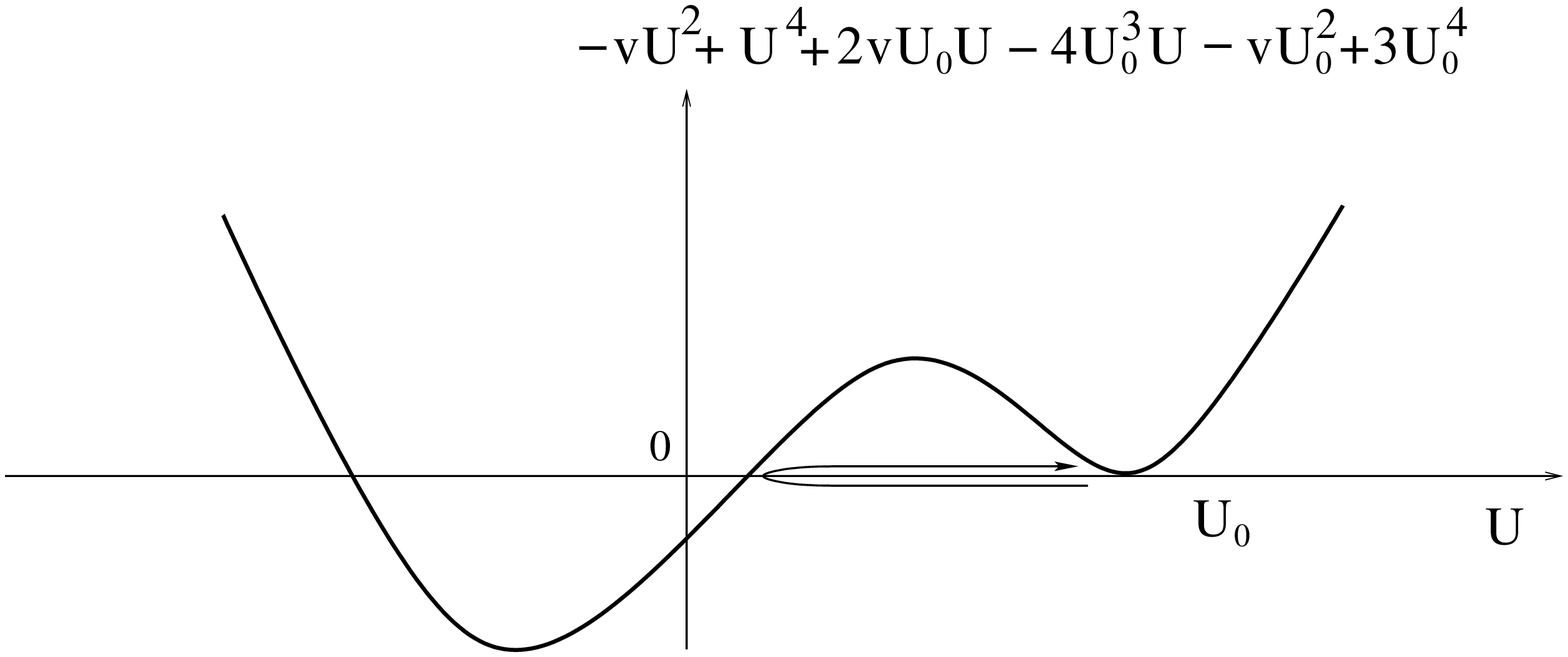}
\end{center}
\caption{The path of integration w.r.t. $U$ corresponding to
a one-soliton solution on a "pedestal" for equation
(\ref{MinusmKdV}).}
\label{MinusSolPed}
\end{figure}

It is not difficult to see that equation (\ref{MinusmKdV}) does
not have real soliton solutions in ordinary sense and only the
soliton solutions on "pedestal" can exist in this case. The
one-soliton solutions on "pedestal" are defined by the analytic 
formula

$$\pm \,\, \int
{d U \over \sqrt{- v U^{2} + U^{4} + 2 v U_{0} U - 4 U_{0}^{3} U
- v U_{0}^{2} + 3 U_{0}^{4}}} \,\, = \,\, X \, + \, C $$
where the path of integration w.r.t. $U$ is shown at Fig.
\ref{MinusSolPed}.

\begin{figure}
\begin{center}
\includegraphics[width=14.0cm,height=6cm]{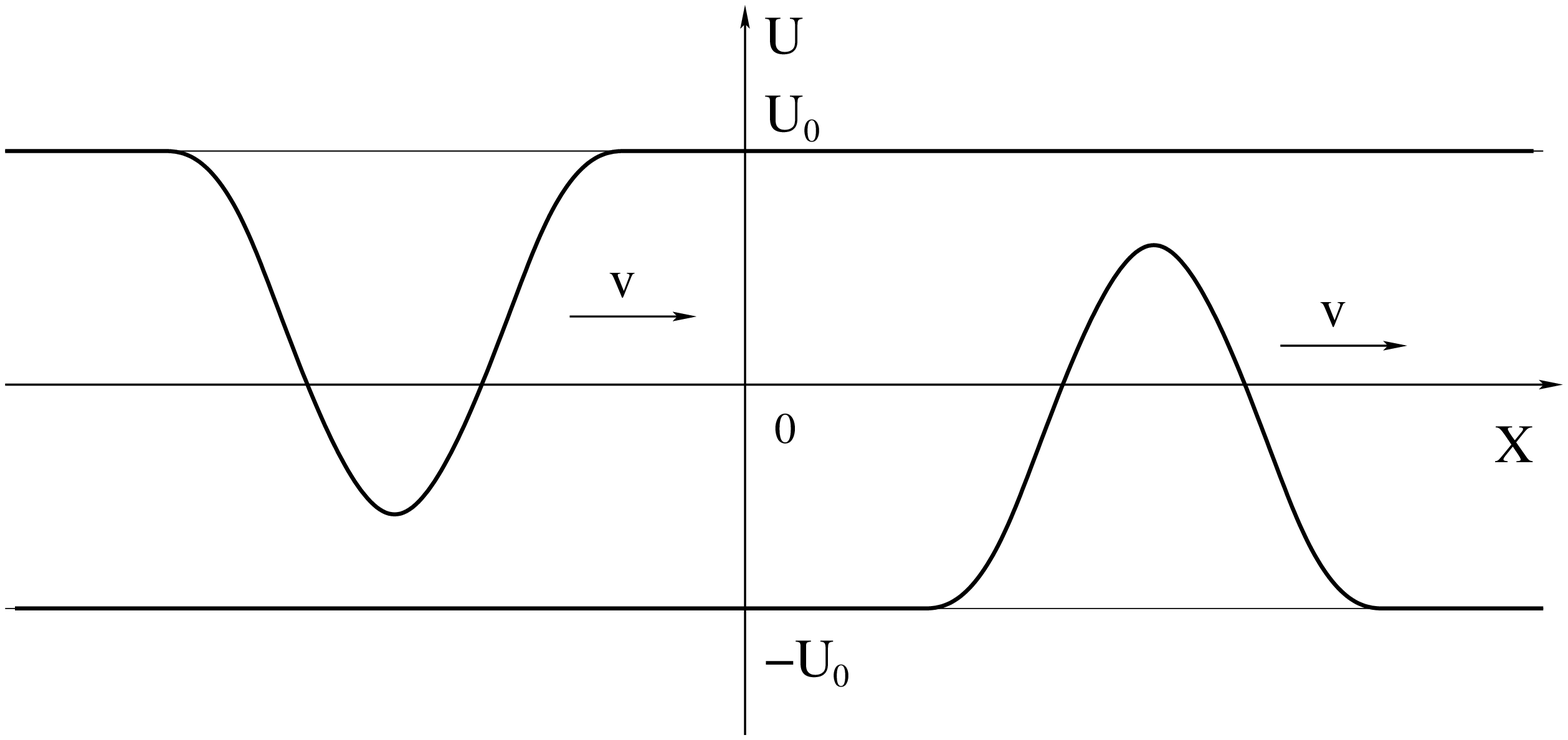}
\end{center}
\caption{The solitons of the hole type and of the particle
type on the positive and negative "pedestals" for equation
(\ref{MinusmKdV}).}
\label{MinusParHolPed}
\end{figure}

Explicit formula for the soliton solution can be written in the
form

$$U \,\, = \,\, \pm \, \left[ U_{0} \,\, - \,\,
{2 a^{2} \over \sqrt{U_{0}^{2} - a^{2}} \,\,
{\rm ch} \, (2 a X - (12 U_{0}^{2} a - 8 a^{3}) T + c_{0})
\, + \, U_{0}} \right] $$
such that the soliton is of the hole-type for the positive "pedestal"
and is of the particle-type for the negative pedestal ($U_{0} > 0$)
(Fig. \ref{MinusParHolPed}).

 The amplitude of soliton

$$A \,\, = \,\, {2 a^{2} \over \sqrt{U_{0}^{2} - a^{2}} +  U_{0}} $$
does not exceed the value $2 U_{0}$ and can be arbitrarily small for
$a \rightarrow 0$. The inverse scattering method and construction
of the $N$-soliton solutions on "pedestal" for equation 
(\ref{MinusmKdV}) were considered in \cite{Romanova} and equation
(\ref{MinusmKdV}) demonstrates that integrable properties are analogous
to those of the KdV equation.

We can see then that equation (\ref{MinusmKdV}) gives a satisfactory
limit of the regimes described by equation (\ref{minus})
($\mu_{0}^{2} < (J_{\perp} - 7 J_{z}) / 7 (J_{\perp} - J_{z})$,
$J_{\perp} > 7 J_{z}$) in the limit $\mu_{0} \rightarrow 0$. 
We have to note, however, that the amplitude of solitons is 
restricted now by the value $2 \mu_{0}$ for $\mu_{0} \rightarrow 0$
and soliton solutions disappear for $\mu_{0} = 0$. Thus,
generation of solitons in the regimes corresponding to equations
(\ref{minus}), (\ref{MinusmKdV}) should be suppressed in the limit
$\mu_{0} \rightarrow 0$. This should be contrasted to the regimes
corresponding to equations (\ref{plus}), (\ref{PlusmKdV}).

\subsection{Appendix. Step decay close to half-filling}

\begin{figure}
\begin{center}
\includegraphics[width=14.0cm,height=6cm]{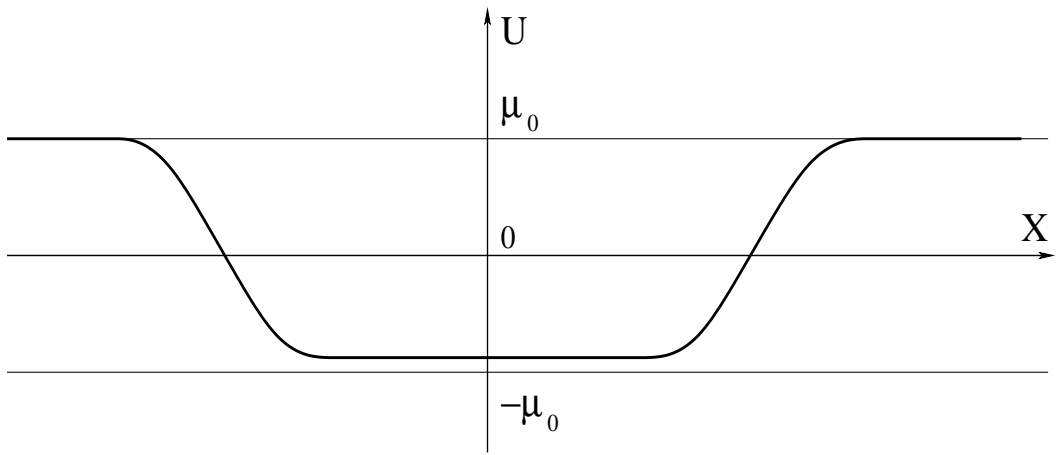}
\end{center}
\caption{The limiting form of a soliton solution for equation
(\ref{MinusmKdV}).}
\label{LimSol}
\end{figure}

One can use  the inverse scattering method to solve
initial value problems with localized initial perturbations for
equations (\ref{MinusmKdV}) or (\ref{PlusmKdV}) very similarly
to what we discussed for equations (\ref{minus}) or (\ref{plus}). 
However, 
localized initial perturbation ($U (X) \rightarrow 0$, 
$X \rightarrow \pm \infty$) will be a source of solitons at final 
stages only for equation (\ref{PlusmKdV}) for
$\mu_{0} = 0$. The soliton part will be absent in the solutions
of (\ref{MinusmKdV}). We also point out that for  small
$\mu_{0} \neq 0$ and big amplitude of initial perturbation for
equation (\ref{MinusmKdV}) ($V_{0} >> \mu_{0}$) the "limiting"
soliton (Fig. \ref{LimSol}) in the limit $T \rightarrow \infty$
can arise (\cite{PerFridYel}).

We also discuss briefly dynamics starting from 
the step-like initial state for
equations (\ref{MinusmKdV}) and (\ref{PlusmKdV}) and the 
asymptotes of the corresponding solutions for $T \rightarrow \infty$.
According to the type of the solutions we considered above we
will consider now the initial data such that

$$U (X) \, \rightarrow \, U_{1} \,\,\, , \,\,\,
X \rightarrow - \infty \,\,\,\,\,\,\,\, , \,\,\,\,\,\,\,\,
U (X) \, \rightarrow \, U_{2} \,\,\, , \,\,\,
X \rightarrow + \infty $$
where both $U_{1}$ and $U_{2}$ are supposed to be small.

 Let us note first of all that the situation here is not pretty
much different from those shown at Fig. \ref{StepOscMinus} and
Fig. \ref{StepOscPlus} in the case when $U_{1}$ and $U_{2}$
have the same signs (say $U_{1}, U_{2} > 0$). So, the new
features will arise here only in the case of different signs
of $U_{1}$ and $U_{2}$ both for equations (\ref{MinusmKdV})
and (\ref{PlusmKdV}).

\vspace{0.5cm}

 Let us start again with equation (\ref{PlusmKdV}).

\begin{figure}
\begin{center}
\includegraphics[width=14.0cm,height=12cm]{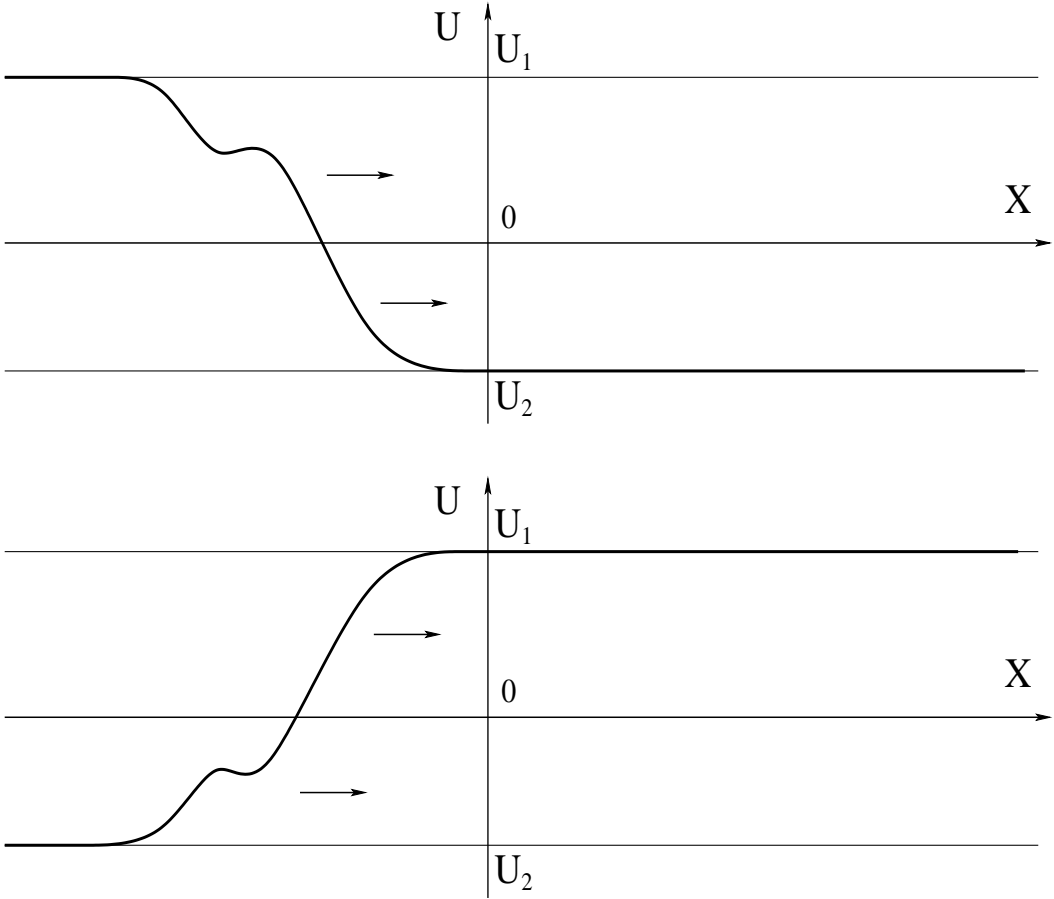}
\end{center}
\caption{Appearance of oscillations for two different 
kinds of steps for equation (\ref{mKdV}) in the case of 
different signs of $U_{1}$ and $U_{2}$.}
\label{mu1mu2Osc}
\end{figure}

 We have to say first that the oscillation region arises now
for the both kinds of steps for the different signs of
$U_{1}$ and $U_{2}$ (see Fig. \ref{mu1mu2Osc}) and the 
situation with just a decreasing of the steepness of initial data
shown at Fig. \ref{r1r2Hydr} is impossible in this case.

\begin{figure}
\begin{center}
\includegraphics[width=14.0cm,height=12cm]{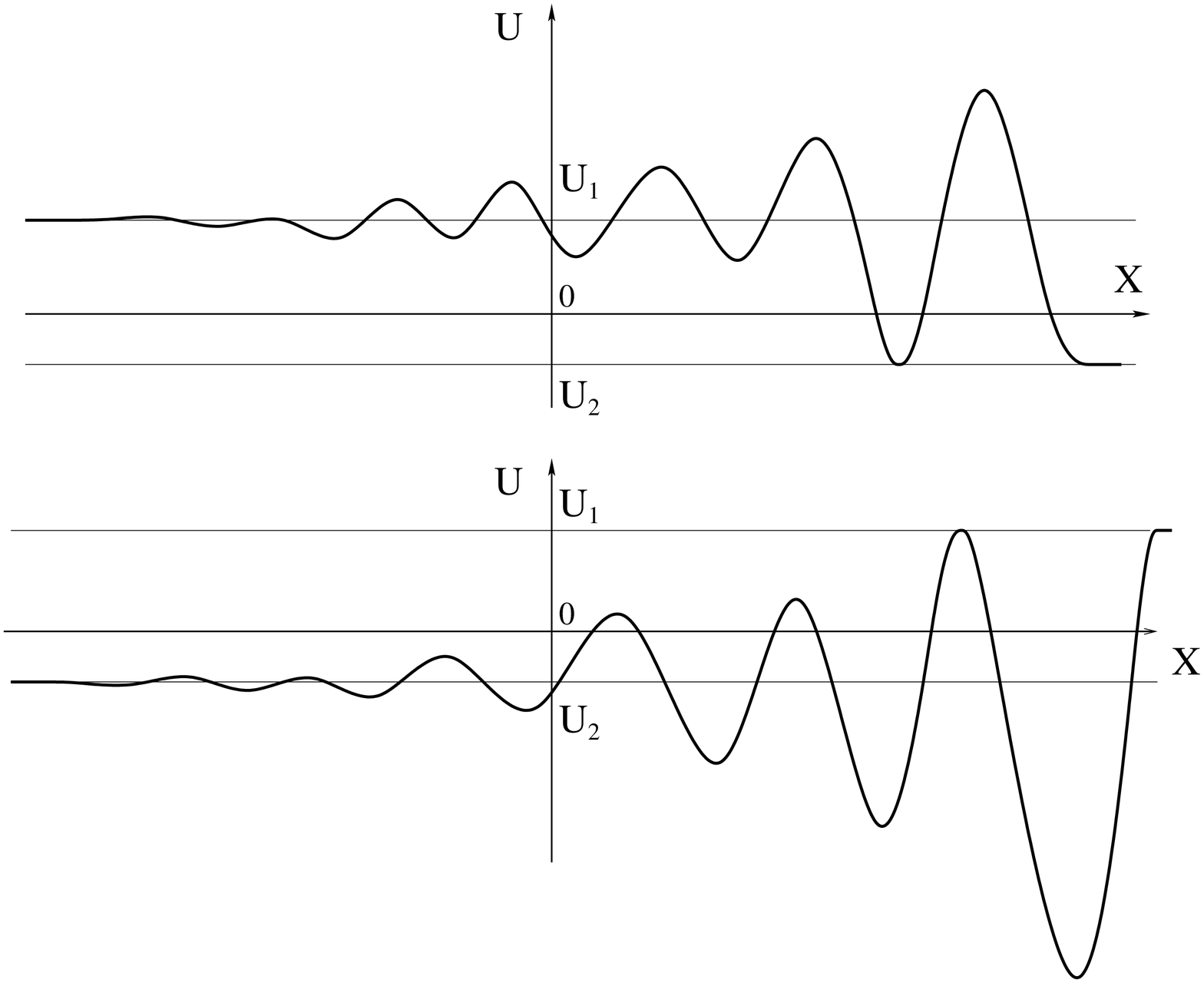}
\end{center}
\caption{The particle-type and hole-type solitons on the
pedestals, $U_{1}$ and $U_{2}$, arising for two types of the
step-like initial data for equation (\ref{PlusmKdV}).}
\label{mu1mu2SolPlus}
\end{figure}

 Both the situations shown at Fig. \ref{mu1mu2Osc} for 
(\ref{PlusmKdV}) result in the generation of solitons on the
final stage which have the particle type in the first and the
hole type in the second situation (Fig. \ref{mu1mu2SolPlus}).

 We can see that the regimes of decay of step-like initial data
for (\ref{PlusmKdV}) include both the regimes coming from 
$\mu_{0} > 0$ and $\mu_{0} < 0$ which is rather natural and
gives a good limit for $\mu_{0} \rightarrow 0$.

\vspace{0.5cm}

 Let us consider now the situation of equation (\ref{MinusmKdV})
corresponding to the small values of $J_{z}$ and $\mu_{0}$. Let
us consider the initial data shown at the top of 
Fig. \ref{mu1mu2Osc} and suppose first that 
$|U_{1}| > |U_{2}|$. At the situation we describe the final 
stage of the oscillations development looks rather similar to
that shown at Fig. \ref{StepOscMinus} which is rather natural
for the limit $\mu_{0} \rightarrow 0$ in the pattern. However,
the limit $|U_{2}| \rightarrow |U_{1}|$ demonstrates quite
new features here which are connected with the arising of a
new solution for equation (\ref{MinusmKdV}). Indeed, for
$|U_{2}| \rightarrow |U_{1}|$ the solitons arising in the
decay of the step-like initial data have a "limiting" form
(Fig. \ref{LimSol}) which is connected with the separation
of the "shock-wave" solution 

\begin{equation}
\label{UTanH}
U \,\, = \,\, - \, a \,\, {\rm th} \left( a \, X \, - \,
2 \, a^{3} \, T \, + \, c_{0} \right)
\end{equation}
for $|U_{2}| = |U_{1}| = a$. 

 Solution (\ref{UTanH}) plays an important role in the decay of 
the step-like initial data we consider for (\ref{MinusmKdV}) for 
$|U_{2}| \geq |U_{1}|$. Let us say that for general initial 
data having the form

$$U (-\infty) \,\, = \,\, - \, U (+\infty) \,\, = \,\, a $$
all the parts including (\ref{UTanH}), solitons and the
"wave-train" will generically arise (\cite{PerFridYel}).

\begin{figure}
\begin{center}
\includegraphics[width=14.0cm,height=6cm]{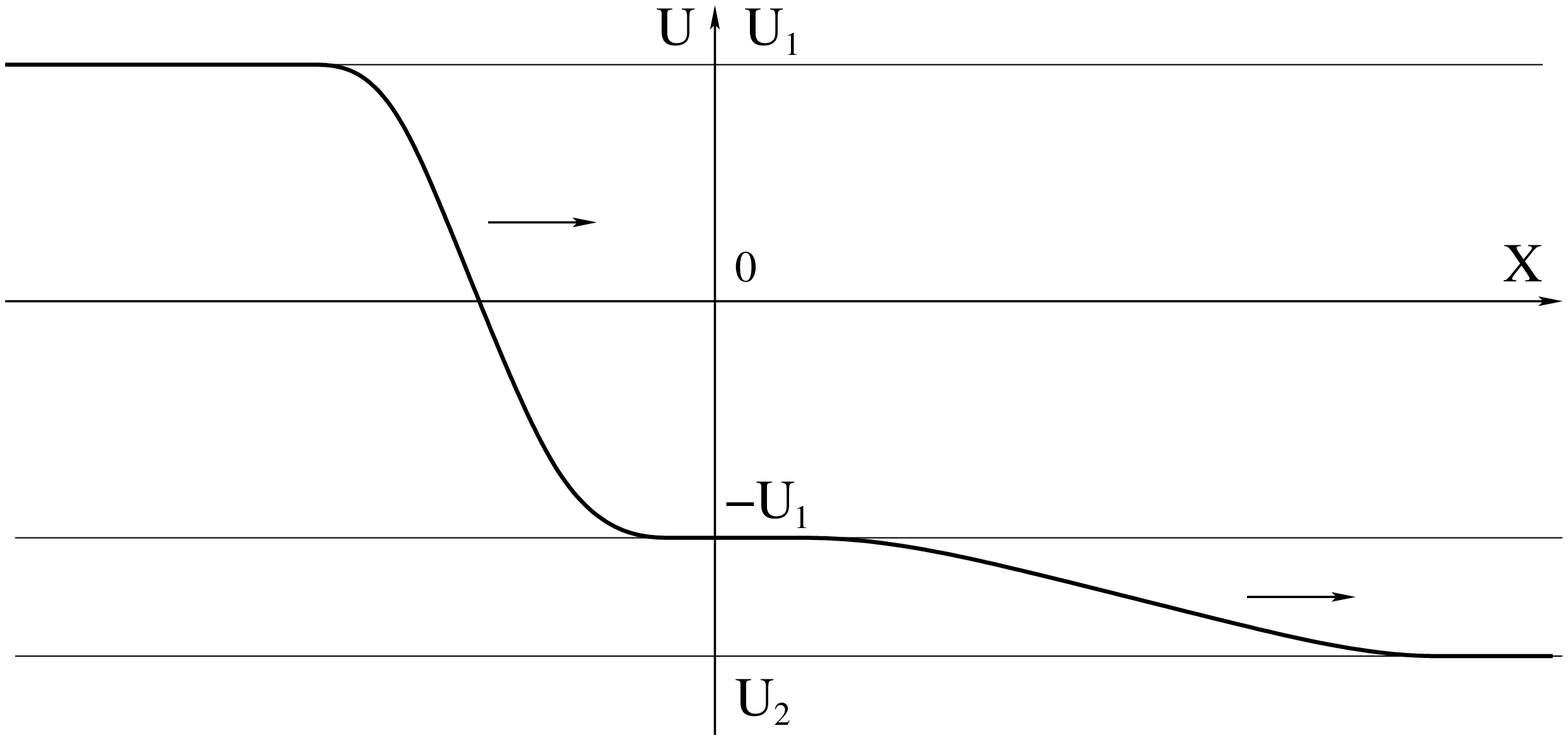}
\end{center}
\caption{Additional step with decreasing steepness
($T \rightarrow \infty$) arising after the separation of solution 
(\ref{UTanH}) for $|U_{2}| > |U_{1}|$.}
\label{AdditStep}
\end{figure}

 It's not difficult to understand also that for 
$|U_{2}| > |U_{1}|$ an additional step of the height
$|U_{2}| - |U_{1}|$ with the decreasing steepness
will arise near the level $U_{2}$
after the separation of solution (\ref{UTanH})
(Fig. \ref{AdditStep}).

 We have to say now that the step-like initial conditions of the
second type (the bottom of Fig. \ref{mu1mu2Osc}) can be investigated
just by the change $U \rightarrow - U$.

\begin{figure}
\begin{center}
\includegraphics[width=14.0cm,height=12cm]{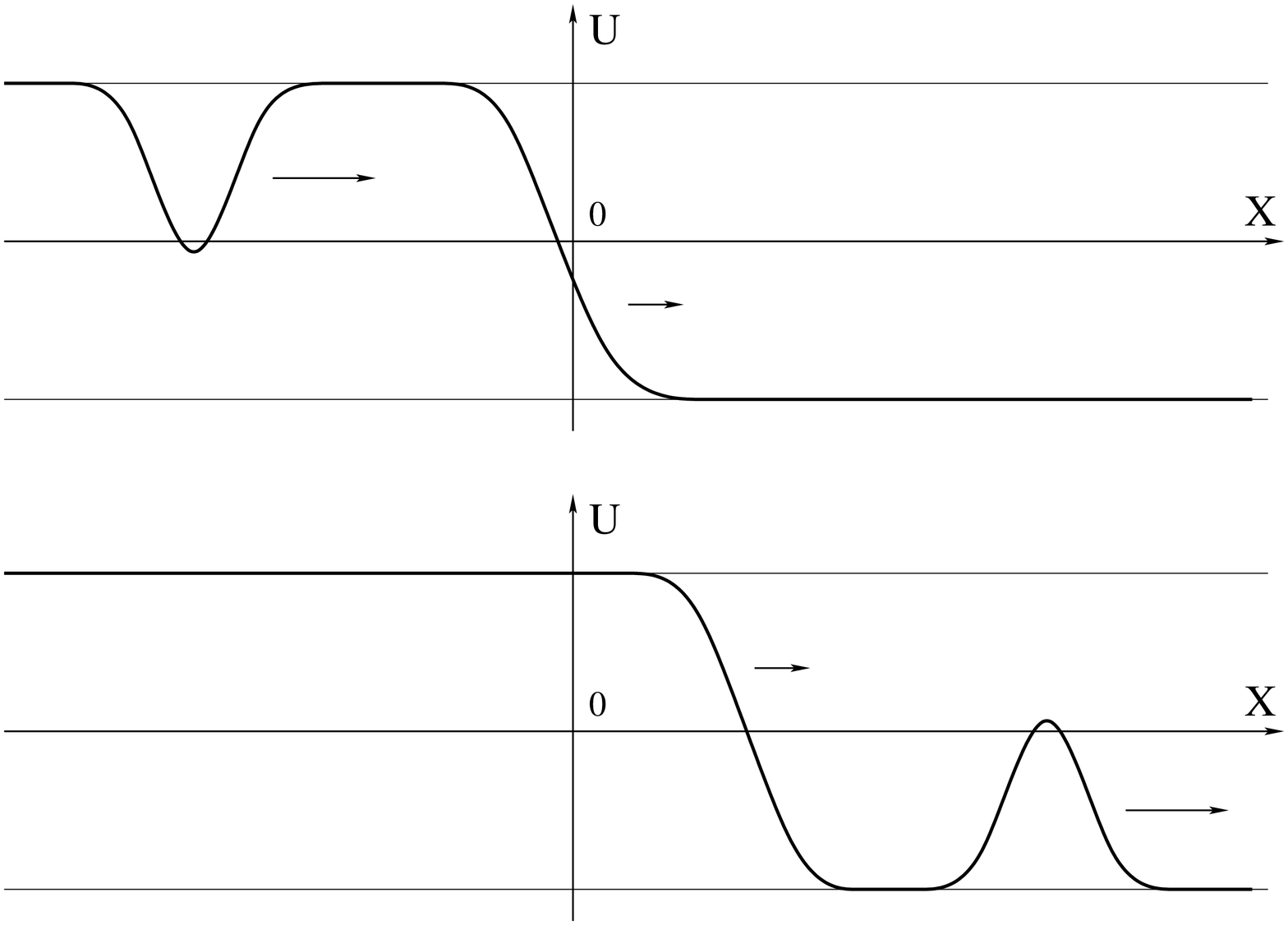}
\end{center}
\caption{Soliton "flip" after interacting with the
"shock-wave" solution (\ref{UTanH}) for equation
(\ref{MinusmKdV}).}
\label{SolFlip}
\end{figure}

 Let us mention here also the very interesting solutions of
(\ref{MinusmKdV}) including the soliton part and solution
(\ref{UTanH}). The soliton solutions coexist with solution
(\ref{UTanH}) and the interaction of a soliton with (\ref{UTanH})
results in the phase shift and the soliton "flip"
(Fig. \ref{SolFlip}).

\begin{figure}
\begin{center}
\includegraphics[width=14.0cm,height=7cm]{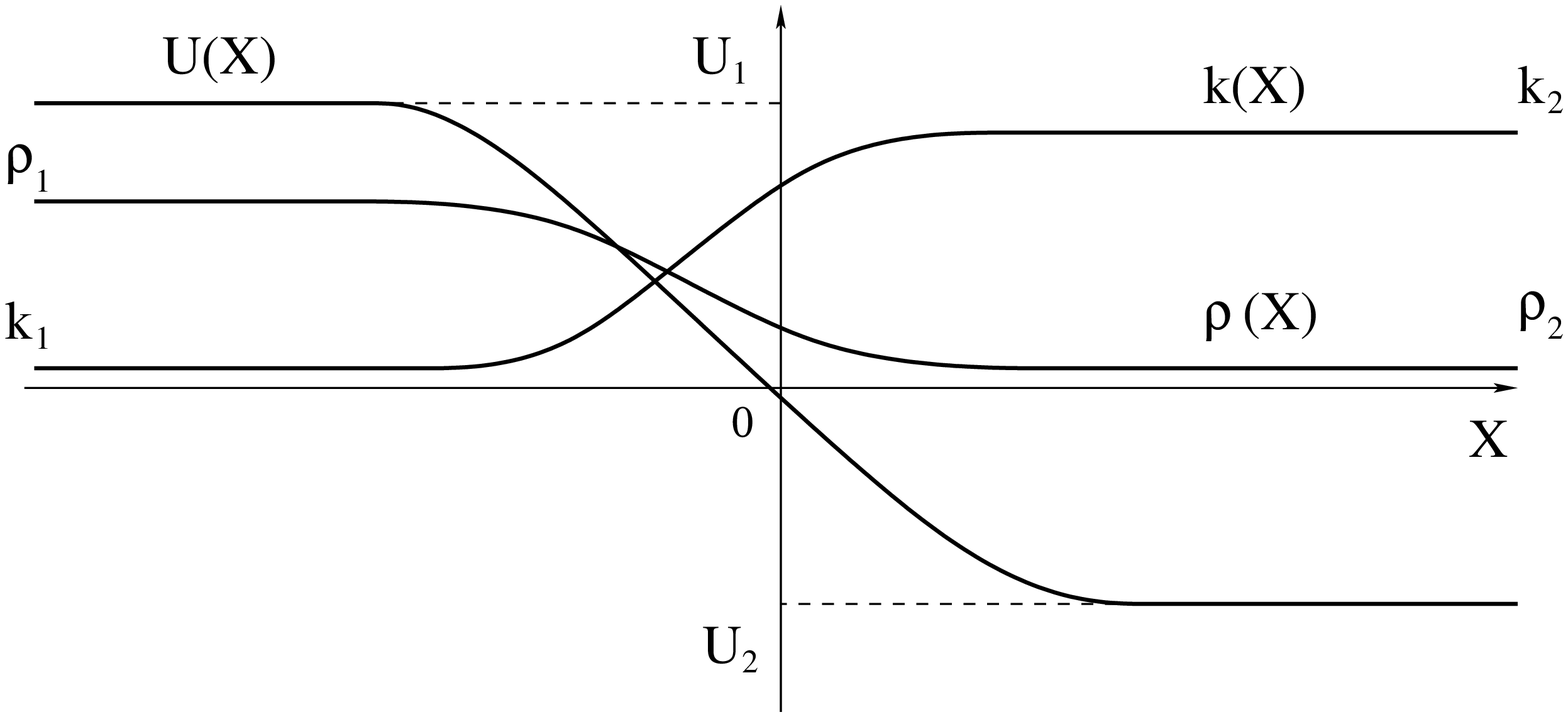}
\end{center}
\caption{Correspondence between functions $U (X)$ and
$(\rho (X), k (X))$ in the left-moving part of the step-like
initial conditions for $T > 0$.}
\label{Urhok}
\end{figure}

\begin{figure} 
\begin{center}
\includegraphics[width=14.0cm,height=7cm]{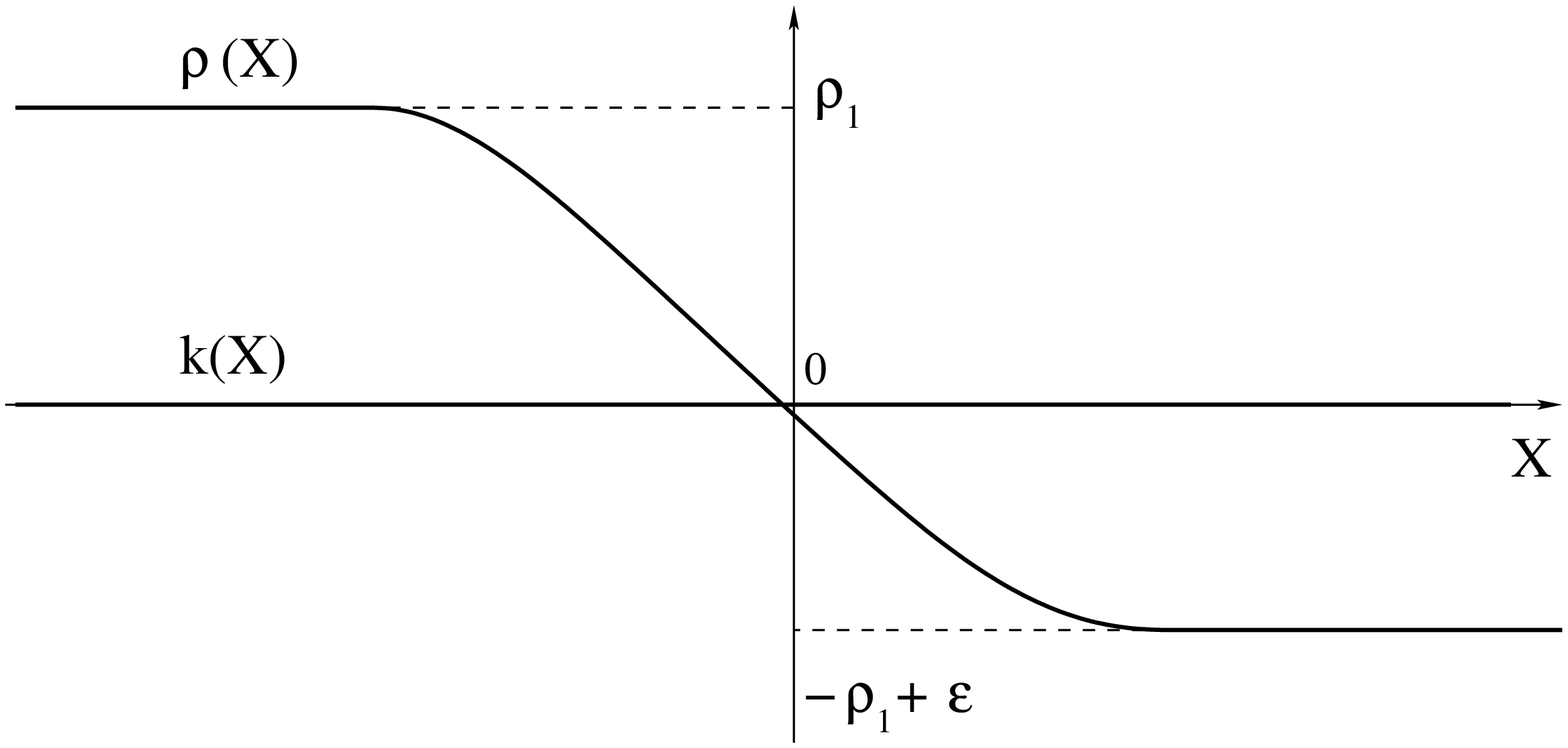}  
\end{center}
\caption{Initial conditions for $\rho (X)$ and $k (X)$, which give rise
to the case $U_{1} \sim - U_{2}$, $|U_{1}| > |U_{2}|$, after 
separation of the left- and right-moving parts for $T > 0$.}
\label{SkewSymmrho}
\end{figure}

 Finally, we point out again that while considerations above were
given for the function $U (X)$, representing Riemann invariants
$r^{\{1,2\}} (X)$, we can express the results in terms of physical
variables $\rho$ and $k$ using equation (\ref{rhokExp}) (we also
remind the readers that our analysis assumes the limit $\rho \rightarrow 0$
and $k \rightarrow 0$). We find that for $|U_{1}| > |U_{2}|$ (Fig. \ref{Urhok})
the case
$$U_{1} \sim - U_{2}$$
corresponds to the case $\rho (X) > 0$, $k (X) > 0$. 
For $|U_{1}| - |U_{2}| \ll |U_{1}|$ we  also have
$\rho_{2} \ll \rho_{1}$, $k_{1} \ll k_{2}$ (see Fig. \ref{Urhok}).
It is not difficult to see that conditions of this type can
arise naturally after separating the right- and 
left-moving parts of initial conditions, as shown in 
Fig. \ref{SkewSymmrho} .

\vspace{0.5cm}

\bibliographystyle{plain}
\bibliography{references_solitons}

\begin{thebibliography}{99}

\bibitem{AblSeg} M.J. Ablowitz, H. Segur.
Solitons and the Inverse Scattering Transform.
Society for Industrial and Applied Mathematics (SIAM), 
Philadelphia, 1981. 

\bibitem{Ahufinger2004} Ahufinger, V.  and Sanpera, A.  and Pedri, P.  and 
Santos, L.  and Lewenstein, M., Creation and mobility of discrete solitons 
in Bose-Einstein condensates, Phys. Rev. A {\bf 69} : 5 (2004), 053604.

\bibitem{Ahufinger2005} Ahufinger, V.  and Sanpera, A.,
Lattice Solitons in Quasicondensates, Phys. Rev. Lett. {\bf 94} : 13
(2005), 130403.

\bibitem{Altman2005} Altman, E.  and Polkovnikov, A.  and Demler, E.  and 
Halperin, B. I. and Lukin, M. D., Superfluid-Insulator Transition in a 
Moving System of Interacting Bosons, Phys. Rev. Lett. {\bf 95} : 2
(2005), 020402.

\bibitem{Bakr2010} W. Bakr et al., Nature {\bf 462} (2009), 74.

\bibitem{BenFair} Benjamin, T.B. and Feir, J.E., 
The disintegration of wave trains on deep water. Part 1. Theory. 
J. Fluid Mech. 27 (1967) 417-430. Feir, J.E. Discussion: 
Some results from wave pulse experiments, 
Proc. R. Soc. Lond. A 299 (1967) 54-58.


\bibitem{BespTal} V. I. Bespalov and V. I. Talanov. JETP Letters.
{\bf 3} (1966), 307.












\bibitem{BetAbWieg1} E. Bettelheim, A. G. Abanov, P. Wiegmann.
Orthogonality catastrophe and shock waves in a non-equilibrium Fermi gas.
arXiv:cond-mat/0607453 , Phys. Rev. Lett. 97, 246402 (2006).  

\bibitem{BetAbWieg2} E. Bettelheim, A. G. Abanov, P. Wiegmann.  
Quantum Shock Waves - the case for non-linear effects in dynamics of 
electronic liquids. arXiv:cond-mat/0606778 , 
Phys.Rev.Lett. 97 (2006) 246401.

\bibitem{BetAbWieg3} E. Bettelheim, A. G. Abanov, P. Wiegmann.
Nonlinear Dynamics of Quantum Systems and Soliton Theory.
arXiv:nlin/0605006 , J.Phys. A40 (2007) F193-F208 .

\bibitem{BIMMZ} Borgad L.A., Its A.R., Matveev V.B., Manakov S.V.,
Zakharov V.E. Phys. Lett. {bf 63 A}, N 3 (1979) p. 205. 

\bibitem{Bloch2005} I. Bloch, Nature Physics {\bf 1} (2005), 23.

\bibitem{Bloch2008_1} Bloch, Immanuel  and Dalibard, Jean  and Zwerger, 
Wilhelm, Many-body physics with ultracold gases, Rev. Mod. Phys.
{\bf 80} : 3 (2008), 885--964.

\bibitem{Bloch2008_2} I. Bloch, Science {\bf 319} (2008), 1202.

\bibitem{Burger1999} Burger, S.  and Bongs, K.  and Dettmer, S.  and 
Ertmer, W.  and Sengstock, K.  and Sanpera, A.  and Shlyapnikov, G. V. and 
Lewenstein, M., Dark Solitons in Bose-Einstein Condensates, Phys. Rev. 
Lett. {\bf 83} : 25 (1999), 5198--5201.
 
\bibitem{Castin2009} Y. Castin, Eur. Phys. Journal B {\bf 68} (2007),
556.

\bibitem{Daley2004} A. Daley et al., Journal of Statistical Mechanics: 
Theory and Experiment, {\bf 2004} : 4 (2004), 04005.

\bibitem{Damski2003} Damski, B.  and Zakrzewski, J.  and Santos, L.  and 
Zoller, P.  and Lewenstein, M., Atomic Bose and Anderson Glasses in 
Optical Lattices, Phys. Rev. Lett. {\bf 91} : 8 (2003), 080403.

\bibitem{Denschlag2000} J. Denschlag et al., Science {\bf 287} (2000),
97.

\bibitem{Druma} V.S. Druma, On analytic solution of the two-dimensional
Korteweg-de Vries equation, JETP Lett., {\bf 19} : 12 (1974), 219-225.

\bibitem{Duan2003} Duan, L.-M.  and Demler, E.  and Lukin, M. D.,
Controlling Spin Exchange Interactions of Ultracold Atoms in Optical 
Lattices, Phys. Rev. Lett. {\bf 91} : 9 (2003), 090402.

\bibitem{Eiermann2004} Eiermann, B.  and Anker, Th.  and Albiez, M.  and 
Taglieber, M.  and Treutlein, P.  and Marzlin, K.-P.  and Oberthaler, 
M. K., Bright Bose-Einstein Gap Solitons of Atoms with Repulsive 
Interaction, Phys. Rev. Lett. {\bf 92} : 23 (2004), 230401.

\bibitem{Fallani2004} Fallani, L.  and De Sarlo, L.  and Lye, J. E. and 
Modugno, M.  and Saers, R.  and Fort, C.  and Inguscio, M.,
Observation of Dynamical Instability for a Bose-Einstein Condensate in a 
Moving 1D Optical Lattice, Phys. Rev. Lett. {\bf 93} : 14 (2004),
140406.

\bibitem{FerPasUl} E. Fermi, J.R. Pasta, and S. Ulam.
"Studies of Nonlinear Problems I.", Los Alamos Report No.
LA-1940, 1955.

\bibitem{Fradkin1991} E. Fradkin, Field Theories of Condensed Matter 
Systems, Addison-Wesley Publishing Company (1991).

\bibitem{GelLev} I.M. Gelfand, B.M. Levitan., Izvestia Akad. Nauk 
S.S.S.R., Ser. Math. {\bf 15} (1951), 309.

\bibitem{Gemelke2009} N. Gemelke et al., Nature {\bf 460}
(2009), 995.  

\bibitem{GGKM} G.S. Gardner, J.M. Green, M.D. Kruskal, R.M. Miura.
Phys. Rev. Lett. {\bf 19} , 1095 (1967).

\bibitem{Giamarchi2004} T. Giamarchi, Quantum Physics in One dimension,
Oxford Science Publishing (2004).

\bibitem{Greiner2008} M. Greiner and S. Foelling,
Nature {\bf 453} (2008), 736.

\bibitem{Greiter1989} M. Greiter and F. Wilczek and E. Witten,
Mod. Phys. Lett. {\bf 3} (1989), 405.

\bibitem{GurPit1} A.V. Gurevich, L.P. Pitaevskii.,
Decay of initial discontinuity in the Korteweg - de Vries equation,
{\it JETP Letters} {\bf 17} (1973), 193-195.

\bibitem{GurPit2} A.V. Gurevich, L.P. Pitaevskii.,
Nonstationary structure of a collisionless shock waves,
{\it Sov. Phys. JETP} {\bf 38} (1974), 291-297.

\bibitem{Haller2009} Haller et al., Science {\bf 325} (2009), 1224.

\bibitem{Halperin1969} Halperin, B. I. and Hohenberg, P. C.,
Hydrodynamic Theory of Spin Waves, Phys. Rev. {\bf 188} (1969),
898--918.

\bibitem{Schmiedmayer2010} Heine et al., New J. Phys. {\bf 12} (2010),    
65036.

\bibitem{Hofferberth2007} Hofferberth et al., Nature Physics {\bf 449}
(2007), 324.

\bibitem{Hofferberth2008} Hofferberth et al., Nature Physics {\bf 4}
(2008), 489.

\bibitem{Huber2007} Huber, S. D. and Altman, E.  and B\"uchler, H. P. and 
Blatter, G., Dynamical properties of ultracold bosons in an optical 
lattice, Phys. Rev. B {\bf 75} : 8 (2007), 085106.

\bibitem{Huber2008} Huber, S. D. and Theiler, B.  and Altman, E.  and 
Blatter, G., Amplitude Mode in the Quantum Phase Model, Phys. Rev. Lett.
{\bf 100} : 5 (2008), 050404.

\bibitem{Jackiw1979} R. Jackiw and A. Kerman, Time-dependent variational 
principle and the effective action, Physics Letters A {\bf 71} : 2-3
(1979), 158 - 162.

\bibitem{Jaksch2005} D. Jaksch and P. Zoller,
Annals of Physics {\bf 315} (2005), 52.

\bibitem{Johansson1999} Johansson, Magnus  and Kivshar, Yuri S.,
Discreteness-Induced Oscillatory Instabilities of Dark Solitons,
Phys. Rev. Lett. {\bf 82} : 1 (1999), 85--88.

\bibitem{KadPet} Kadomtsev, B. B., Petviashvili, V. I. "On the 
stability of solitary waves in weakly dispersive media". Sov. Phys. Dokl. 
{\bf 15} (1970), 539-541. 

\bibitem{Karski2009} Karski, M.  and F\"orster, L.  and Choi, J. M. and 
Alt, W.  and Widera, A.  and Meschede, D., Nearest-Neighbor Detection of 
Atoms in a 1D Optical Lattice by Fluorescence Imaging, Phys. Rev. Lett.
{\bf 102} : 5 (2009), 053001.

\bibitem{KayMos} I. Kay, H.E. Moses, Nuovo Cimento {\bf 3} (1956), 276;
J. Appl. Phys. {\bf 27} (1956), 1503.

\bibitem{Kevrekidis2003} Kevrekidis, P. G. and Carretero-Gonz\'alez, R.  
and Theocharis, G.  and Frantzeskakis, D. J. and Malomed, B. A.,
Stability of dark solitons in a Bose-Einstein condensate trapped in an 
optical lattice, Phys. Rev. A {\bf 68} : 3 (2003), 035602.

\bibitem{Khalatnikov1978} I. Khalatnikov and V.V. Lebedev, 
J. Low Temp. Phys. {\bf 32} (1978), 789.

\bibitem{Khaykovich2002} L. Khaykovich et al.,
Science {\bf 287} (2002), 97.

\bibitem{Kinoshita2006} T. Kinoshita et al., Nature {\bf 440} (2006),
900.

\bibitem{Kivshar1989} Kivshar, Yuri S. and Malomed, Boris A.,
Dynamics of solitons in nearly integrable systems, Rev. Mod. Phys.
{\bf 61} : 4 (1989), 763--915.

\bibitem{Kuklov2003} Kuklov, A. B. and Svistunov, B. V.,
Counterflow Superfluidity of Two-Species Ultracold Atoms in a Commensurate 
Optical Lattice, Phys. Rev. Lett. {\bf 90} : 10 (2003), 100401.

\bibitem{Kohl2009} Palzer, Stefan  and Zipkes, Christoph  and Sias, Carlo  
and K\"ohl, Michael, Quantum Transport through a Tonks-Girardeau Gas,
Phys. Rev. Lett. {bf 103} : 15 (2009), 150601.

\bibitem{KrusZab} M.D. Kruskal, N.J. Zabusky. 
Stroboscopic-Perturbation Procedure for Treating a Class of 
Nonlinear Wave Equations. Journ. of Math. Phys. {\bf 5} : 2
(1964), 231 - 244.

\bibitem{Krutitsky2010} K. Krutitsky et al., arXiv:0907.0625
(2010).

\bibitem{Lahaye2009} T. Lahaye et al., Reports on Progress in Physics
{\bf 72} (2007), 126401.

\bibitem{Lancaster2010} Lancaster, Jarrett  and Mitra, Aditi,
Quantum quenches in an  $XXZ$  spin chain from a spatially inhomogeneous 
initial state, Phys. Rev. E {\bf 81} : 6 (2010), 061134.

\bibitem{Lax} P.D. Lax. Comm. Pure Appl. Math. {\bf 21}, 467 
(1968).

\bibitem{LaxLev} P.D. Lax, C.D. Levermore.,
The small dispersion limit for the Korteweg - de Vries equation
I, II, and III. {\it Comm. Pure Appl. Math.}, {\bf 36} (1983),
253-290, 571-593, 809-830.

\bibitem{LaxLevVen} P.D. Lax, C.D. Levermore, S. Venakides.,
The generation and propagation of oscillations in dispersive
IVPs and their limiting behavior, {\it Important developments
in soliton theory 1980-1990}, 205-241, Springer Series in
Nonlinear Dynamics. Springer, Berlin (1993).

\bibitem{Lewenstein2007} M. Lewenstein et al.,
Advances in Physics {\bf 56} (2007), 243.

\bibitem{Lighthill} M.J. Lighthill, Proc. Roy. Soc. {\bf A299}
(1967), 28.

\bibitem{Mar} V.A. Marchenko, Doklady. Akad. Nauk SSSR {\bf 104}
(1955), 695.

\bibitem{MaxRed} T. Maxworthy, L.G. Redekopp., Icarus {\bf 29},
261 (1976).

\bibitem{Miles} Miles J.W. Resonantly interacting solitary waves.
J. Fluid Mech. {\bf 79} (1977), 171-179.

\bibitem{Miura} R.M. Miura. Korteweg - de Vries equation and
generalizations. I. A remarkable explicit nonlinear transformation.
- J. Math. Phys. {\bf 9}, 1202 -1204 (1968).

\bibitem{Mun2007} Mun, Jongchul  and Medley, Patrick  and Campbell, 
Gretchen K. and Marcassa, Luis G. and Pritchard, David E. and Ketterle, 
Wolfgang, Phase Diagram for a Bose-Einstein Condensate Moving in an 
Optical Lattice, Phys. Rev. Lett. {\bf 99} : 15 (2007), 150604.

\bibitem{Murg2007} Murg, V.  and Verstraete, F.  and Cirac, J. I.,
Variational study of hard-core bosons in a two-dimensional optical lattice 
using projected entangled pair states, Phys. Rev. A {\bf 75} : 3
(2007), 033605.

\bibitem{Naumkin1991} Naumkin and Shishmarev, 
The step-decay problem for the Korteweg-de Vries-Burgers equation,
Functional Analysis and Its Applications {\bf 25} : 1 (1991), 16-25.

\bibitem{Nelson2007} K. Nelson and X. Li and D. Weiss,
Nature Phys. {\bf 3} (2007), 556.

\bibitem{Newell} A. C. Newell, Solitons in mathematics and physics.
Society for Industrial and Applied Mathematics (1985).

\bibitem{NovManPitZakh} S.P. Novikov, S.V. Manakov, L.P. Pitaevskii, 
and V.E. Zakharov., Theory of solitons. The inverse scattering
method., Plemun, New York 1984.

\bibitem{Ostrovsky1999} L. Ostrovsky and A. Potapov, 
Modulated Waves. Theory and applications., The Johns Hopkins University 
Press (1999).

\bibitem{Ott2008} H. Ott et al., Nature Phys. {\bf 4} (2008), 949.

\bibitem{PerFridYel} T.L. Perelman, A. Kh. Fridman, M.M. Yelyashevich.
Modified Korteweq - de Vries equation in electrohydrodynamics.
Sov. Phys. JETP. {\bf 39} (1974a), 643-646.

\bibitem{Pitaevskii2003} L. Pitaevskii and S. Stringari,
Bose-Einstein Condensation, Oxford Science Publications (2002)

\bibitem{Polkovnikov2005} Polkovnikov, A.  and Altman, E.  and Demler, E.  
and Halperin, B.  and Lukin, M. D., Decay of superfluid currents in a 
moving system of strongly interacting bosons, Phys. Rev. A {\bf 71} : 6
(2005), 063613. 

\bibitem{Ritter2007} Ritter, Stephan  and \"Ottl, Anton  and Donner, 
Tobias  and Bourdel, Thomas  and K\"ohl, Michael  and Esslinger, Tilman,
Observing the Formation of Long-Range Order during Bose-Einstein 
Condensation, Phys. Rev. Lett. {\bf 98} : 9 (2007), 090402.

\bibitem{Romanova} N. N. Romanova. N-Soliton solution on a pedestal 
of the modified Korteweg-de Vries equation. - Theor. and Math. Phys.
{\bf 39} : 2 (1979), 415-421.

\bibitem{Satsuma} Satsuma J. $N$-soliton solution of the
two-dimensional Korteweg - de Vries equation. J. Phys. Soc.
Japan. {\bf 40} (1976), 286-290.

\bibitem{Sadler2006} L. Sadler et al., Nature {\bf 443} (2006),
312.

\bibitem{Scalettar1995} Scalettar, R. T. and Batrouni, G. G. 
and Kampf, A. P. and Zimanyi, G. T., Simultaneous diagonal and 
off-diagonal order in the Bose-Hubbard Hamiltonian, Phys. Rev. B
{\bf 51} : 13 (1995), 8467--8480.

\bibitem{Scarola2005} Scarola, V. W. and Das Sarma, S.,
Quantum Phases of the Extended Bose-Hubbard Hamiltonian: Possibility of a 
Supersolid State of Cold Atoms in Optical Lattices, Phys. Rev. Lett.
{\bf 95} : 3 (2005), 033003.

\bibitem{Schneider2010} U. Schneider et al., Breakdown of diffusion: From 
collisional hydrodynamics to a continuous quantum walk in a homogeneous 
Hubbard model, arXiv:1005.3545,  2010.

\bibitem{Schmid2002} Schmid, Guido  and Todo, Synge  and Troyer, Matthias  
and Dorneich, Ansgar, Finite-Temperature Phase Diagram of Hard-Core Bosons 
in Two Dimensions, Phys. Rev. Lett. {\bf 88} : 16 (2002), 167208.

\bibitem{Schollwock2005} Schollw\"ock, U., The density-matrix 
renormalization group, Rev. Mod. Phys. {\bf 77} : 1 (2005), 259--315.

\bibitem{Kuhr2010} Sherson et al., Nature {\bf 467} (2010), 68.

\bibitem{Strohmaier2010} Strohmaier, Niels  and Greif, Daniel  and 
J\"ordens, Robert  and Tarruell, Leticia  and
Moritz, Henning  and Esslinger, Tilman  and Sensarma, Rajdeep  and Pekker, 
David  and Altman, Ehud  and Demler, Eugene.,
Phys. Rev. Lett. {\bf 104} : 8 (2010), 080401.

\bibitem{Sutherland2004} B. Sutherland, Beautiful Models,
World Scientific (2004).

\bibitem{Trombettoni2001} Trombettoni, Andrea  and Smerzi, Augusto,
Discrete Solitons and Breathers with Dilute Bose-Einstein Condensates,
Phys. Rev. Lett. {\bf 86} : 11 (2001), 2353--2356.

\bibitem{Trotzky2008} S. Trotzky et al., Science {\bf 319} (2008),
295.

\bibitem{Wadati} M. Wadati. J. Phys. Soc. Japan {\bf 34},
1289 (1973).

\bibitem{Wen1995} X.G. Wen, Advances in Physics {\bf 44}
(1995), 405. 

\bibitem{Wu2001} Wu, Biao  and Niu, Qian, Landau and dynamical 
instabilities of the superflow of Bose-Einstein condensates in optical 
lattices, Phys. Rev. A {\bf 64} : 6 (2001), 061603.

\bibitem{Yulin2003} Yulin, Alexey V. and Skryabin, Dmitry V.,
Out-of-gap Bose-Einstein solitons in optical lattices,
Phys. Rev. A {\bf 67} : 2 (2003), 023611. 

\bibitem{Zab} N.J. Zabusky. Phenomena Associated with the
oscillations of a Nonlinear Model String. In Proceedings of the
Conference on Mathematical Models in the Physical Sciences,
edite by Stefan Drobot (Prentice - Hall, Inc. New York, 1963),
p. 99.

\bibitem{ZabKrus2} N.J. Zabusky and M.D. Kruskal. Interaction of
"solitons" in a collisionless plasma and the recurrence of
initial states. Phys. Rev. Lett. {\bf 15} : 6 (1965), 240-243.

\bibitem{ZakhMan} V.E. Zakharov, S.V. Manakov., Asymptotic
behavior of nonlinear wave systems integrated by the inverse
scattering method., Sov. Phys. JETP {\bf 44} (1) (1976), 106-112.

\bibitem{ZakhShab} V. E. Zakharov and A. B. Shabat, Interaction  
between solitons in a stable medium. {\it Sov. Phys. JETP} 
{\bf 37}, 823-828 (1973).

\bibitem{ZakhShab2} V. E. Zakharov and A. B. Shabat, Integration 
method of nonlinear equations of mathematical physics with the help of 
the inverse scattering problem, Funk. Anal Pril., {\bf 8} : 3 (1974), 
43-53.

\bibitem{Zakrzewski2005} Zakrzewski, Jakub, Mean-field dynamics of the 
superfluid-insulator phase transition in a gas of ultracold atoms,
Phys. Rev. A {\bf 71} : 4 (2005), 043601.

\bibitem{Moritz2010} Zimmermann et al., arXiv:1011.1004 (2010).






\end{thebibliography}

\end{document}